\documentclass[onecolumn,fleqn,usenatbib]{mnras}

\usepackage{newtxtext,newtxmath}
\usepackage[T1]{fontenc}

\DeclareRobustCommand{\VAN}[3]{#2}
\let\VANthebibliography\thebibliography
\def\thebibliography{\DeclareRobustCommand{\VAN}[3]{##3}\VANthebibliography}

\usepackage{graphicx}	% Including figure files
\usepackage{amsmath}	% Advanced maths commands
\usepackage{threeparttable}
\usepackage{hyperref}

\title[Flares on a contact binary]{Optical Flares Detected on a Contact Binary: The First Photometric and Spectroscopic Analysis of a Long-period Low Mass Ratio Contact Binary HAT 307-0007476}

\author[Li et al.]{
Ling-Zhi Li,$^{1}$\thanks{This paper is dedicated to Lingzhi Li in heaven.}
Kai Li,$^{1}$\thanks{corresponding author: kaili@sdu.edu.cn}
Xiang Gao,$^{1}$
Xiao-Dian Chen,$^{2}$
Shuai Feng, $^{1}$
Dong-Yang Gao, $^{1}$
Di-Fu Guo, $^{1}$
Xu Chen, $^{1}$
\\
\newauthor
Xing Gao$^{3}$
Guo-You Sun$^{4}$
Shahidin Yaqup$^{3}$
Chunhai Bai$^{3}$
and Ali Esamdin$^{3}$
\\
$^{1}$ Shandong Key Laboratory of Optical Astronomy and Solar-Terrestrial Environment, School of Space Science and Technology, Institute of Space Sciences,\\ 
Shandong University, Weihai, Shandong 264209, China\\
$^{2}$ CAS Key Laboratory of Optical Astronomy, National Astronomical Observatories, Chinese Academy of Sciences, Beijing 100101, People's Republic of China\\
$^{3}$ Xinjiang Astronomical Observatory, Chinese Academy of Sciences, 150 Science 1-Street, Urumqi 830011, China\\
$^{4}$ Xingming Observatory, Urumqi 830011, China
}

\date{Accepted XXX. Received YYY; in original form ZZZ}

\pubyear{2025}

\begin{document}
\label{firstpage}
\pagerange{\pageref{firstpage}--\pageref{lastpage}}
\maketitle

% Abstract of the paper
\begin{abstract}
This paper presents the photometric and spectroscopic analysis of a long-period totally eclipsing contact binary (HAT 307-0007476) for the first time. This system is a low mass ratio ($q\sim0.114$) and medium contact binary ($f\sim37.1\%$). Two flare events were detected in multiple bands observations in December 2022. The interval between the two flare events is 4 days. The average duration of these two flares is about 2289s. Both the two flares achieve the energy levels of superflares. The excess emission of the H$_\alpha$ line in the LAMOST spectra of this object was analyzed, indicating its chromospheric activity. The $O-C$ diagram showed a long-term orbital period increase, which is due to the mass transfer between the two component stars. We conclude that HAT 307-0007476 is currently in a stable region based on both $J_{spin}/J_{orb}$ and the comparison between the instability parameters and its current values.
\end{abstract}

% Select between one and six entries from the list of approved keywords.
% Don't make up new ones.
\begin{keywords}
{ binaries: eclipsing; stars: activity; stars: flare; stars: individual:}
\end{keywords}

%%%%%%%%%%%%%%%%%%%%%%%%%%%%%%%%%%%%%%%%%%%%%%%%%%

%%%%%%%%%%%%%%%%% BODY OF PAPER %%%%%%%%%%%%%%%%%%

\section{Introduction}
More than 50$\%$ of the stars in the universe are located in binary systems, and contact binaries account for about 95$\%$ of the eclipsing binaries near our Solar system \citep{1948HarAC.960....2R}, this provides a rich natural laboratory for us to investigate the stellar evolution. Decades of research on contact binaries has led to significant achievements in understanding stellar formation, and evolution, and advancing astrophysics.numerous intriguing and rewarding topics within the study of contact binaries, such as the short period cut-off, the existence of the low mass ratio limit, flare activity.

Theoretical studies indicate the presence of a lower limit for the mass ratio of contact binaries, with $q_{min}$ ranging from approximately 0.05 to 0.09 \citep{2006MNRAS.369.2001L, 2009MNRAS.394..501A, 2010MNRAS.405.2485J}. Once the mass ratio reaches the critical threshold where the rotational angular momentum of the tidally locked component exceeds one-third of its orbital momentum, the Darwinian instability \citep{1893Obs....16..172D} leads to a rapid decrease in angular momentum due to tidal dissipation and non-conservative mass loss via the second Lagrange point. Subsequently, the contact binary would merge, resulting in a single, rapidly rotating star. However, observations have revealed contact binaries with mass ratios that fall below the theoretically predicted lower limit. Notable examples include TYC 3801-1529-1 with a mass ratio of 0.036 \citep{2024A&A...692L...4L}, V1187 Her with a mass ratio of 0.044 \citep{2019PASP..131e4203C},  TYC 4002-2628-1 with a mass ratio of 0.048 \citep{2022MNRAS.517.1928G}. Identifying contact binaries that have mass ratios falling below the theoretical lower limit presents a significant challenge to current theoretical models. Therefore, analyzing contact binaries with mass ratios around or less than 0.1 will be crucial for refining our understanding of the low mass ratio cutoff \citep{2022AJ....164..202L,2023MNRAS.519.5760L}. 

The studies of mass ratio and orbital period of contact binaries are useful for understanding the contact binaries evolution. \cite{2019AAS...23344805M,2022AAS...24030801M} proposed angular momentum conserving models to explain the contact binaries evolution, illustrating the relation between the orbital period and the lower limit of the mass ratio. \cite{2022ApJS..262...12K} plotted the mass ratio versus orbital period for the contact binaries based on the models by \cite{2019AAS...23344805M,2022AAS...24030801M}. \cite{2022ApJS..262...12K} also found that the instability mass ratio increases from 0.045 to 0.15 with increasing P, and proposed that merger candidates are those systems with long periods and mass ratios near the lower limit of the mass ratio. Long-period contact binaries (P$\textgreater$0.5d) are rare \citep{2019AAS...23344805M,2022AAS...24030801M} and the evolutionary mechanism of long-period contact binaries has yet to be characterized. So, more studies on long-period low mass ratio contact binaries are useful.

Flare events occur as a result of a violent release of magnetic energy in the outer atmosphere of a stellar at the surface \citep{2002ApJ...577..422S, 2010ARA&A..48..241B, 2010ApJ...714L..98K, 2013ApJS..209....5S}. Our understanding of the flares is mainly derived from our observations of flares on the Sun, where we can observe them in great detail and at high frequency. The typical energy range for solar flares is about $10^{29}$-$10^{32}$ erg \citep{2002ApJ...577..422S}, and the duration is minutes to hours. In addition, we know that much more energetic flares happen in many kinds of stars \citep{1989ApJ...337..927S}. \cite{2000ApJ...529.1026S} reported nine flares on solar-type stars, with the energy range $10^{33}$-$10^{38}$ erg, and defined flare events $\geq$100 times the maximum solar flare as superflare. \cite{2012Natur.485..478M} detected 365 superflares on solar-type stars with the Kepler data, with the durations of several hours, and amplitudes of order 0.1–1$\%$ of the stellar luminosity. In the follow-up extension, \cite{2013ApJS..209....5S} found 1547 superflares, and they argued that the physical origin of superflares may be attributed to the presence of supermassive star spots.

Astronomers also detect flare events on close binaries. Strong interactions between two components of a binary system are likely to produce magnetic activity \citep{1984ApJ...279..763N}. 
\cite{2012MNRAS.419.1219P} analyzed five RS CVn-type binaries with seven flares. \cite{2016ApJS..224...37G} identified 234 flaring eclipsing binaries in the Kepler Eclipsing Binary Catalog. 
VW Cephei is the first contact binary with flare events that have been observed simultaneously in both the X-ray and radio wavelengths \citep{1988ApJ...330..922V}. \cite{2014ApJS..212....4Q} analyzed 15 optical flares recorded in the i band on the contact binary CSTAR 038663.  \cite{2024MNRAS.527.3982L} used TESS high-cadence data to identify two exceptional flares that are confirmed for the first time to be from the primary component of a contact binary. Contact binaries are an important source for understanding the fundamental physical parameters of stars, and the study of flaring contact binaries helps to understand the relationship between stellar activity and stellar physical parameters, which is a new hot spot in stellar physics at present \citep{2018A&A...616A.108B,2023A&A...674A..30L}.

In this work, we analyzed a totally eclipsing contact binary, HAT 307-0007476, which was first discovered by Hungarian-made Automated Telescope Network (HATNet) \citep{2010MNRAS.408..475H}. It was first classified as a W UMa-type contact binary according to the Catalina Sky Surveys (CSS; \citealt{2014ApJS..213....9D}) periodic variable star catalog. Its orbital period is 0.5329326 days, the mean V band magnitude is 13.83 mag, and the amplitude is 0.26 mag according to All-Sky Automated Survey for Supernovae (ASAS-SN; \citealt{2019MNRAS.486.1907J}) Catalog of Variable Stars. This paper presents the first photometric and spectroscopic investigation for HAT 307-0007476, and reports the flare activities detected on this target.

\section{Observations and Data Reduction} \label{sec:DATA}
\subsection{New Photometry}
The new photometric observations of HAT 307-0007476 were carried out in the V, $R_c$, $I_c$, $g^{'}, r^{'}$ and $i^{'}$ bands with four
telescopes. The details of the telescopes are described as follows: (1) The Weihai Observatory 1.00-m Cassegrain telescope of Shandong University (hereafter WHOT, \citealt{2014RAA....14..719H}). The WHOT is equipped with the PIXIS  $2K\times2K$ camera, its field of views is about $12^{\prime} \times 12^{\prime}$; (2) The 0.60-m Ningbo Bureau of Education and Xinjiang Observatory Telescope (hereafter NEXT). The NEXT is equipped with the FLI PL230-42 $2K\times2K$ camera, its field of views is about $22^{\prime} \times 22^{\prime}$; (3) The 0.85-m reflecting telescope at Xinglong Station of  National Astronomical Observatories (hereafter XL, \citealt{2009RAA.....9..349Z}). The XL was equipped with the Andor $2K\times2K$ CCD camera, its field of views is about $35^{\prime} \times 35^{\prime}$; (4) The Xingming Observatory 0.43-m Photometric Auxiliary Telescope (hereafter PAT). The PAT was equipped with the QHY4040 $4K\times4K$ CCD camera, its field of views is about $43^{\prime} \times 43^{\prime}$. The photometric observation log is listed in Table \ref{table:observation}.
%Table 1
\begin{center}
\begin{table}\centering
\footnotesize
\caption{Time Series CCD Photometric Observations for HAT 307-0007476.}\label{table:observation}
\setlength{\tabcolsep}{4mm}{
\begin{tabular}{cccccccc}
\hline\hline
Date (UT)        & HJD (start)    & HJD (end)  &   Filter\&Errors (mag)  &  Exposure (s) & $N_{obs}$     & Telescope  & Type \\
                       & 2450000.00+              &2450000.00+              &               &                 &     & &\\
\hline                                               

2022 Dec. 19   &9933.11  &9933.28              &$g^{'}0.007/r^{'}0.009/i^{'}0.006$        &  70/60/80     & 53         &  NEXT     & Light curve\\  
2022 Dec. 20   &9934.31  &9934.42             & $g^{'}0.009/r^{'}0.006/i^{'}0.009$      &70/60/80     & 36           &  NEXT   & Light curve \\  
2022 Dec. 21   &9935.05  &9935.26               &$g^{'}0.006/r^{'}0.007/i^{'}0.006$       &70/60/80       & 69           &  NEXT  & Light curve  \\ 
2022 Dec. 25   &9939.11   &9939.32            &$g^{'}0.006/r^{'}0.007/i^{'}0.005$       &70/60/80        & 68          &  NEXT  & Light curve   \\
2023 Nov. 7    &10256.04   &10256.36           & V0.007/$R_c$0.007/$I_c0.007$      &80/50/45       & 121          & WHOT  &  Light curve \\
2023 Nov. 18   &10267.08   &10267.25           & V0.009/$R_c$0.008/$I_c0.006$       &80/50/45       & 47          & WHOT   & Light curve \\
2023 Nov. 22   &10271.36   &10271.46           & $g^{'}0.015/r^{'}0.013/i^{'}0.015$  &70/60/80      &34          & PAT  & Minimum \\
2023 Dec. 5   &10284.03   &10284.20           & V0.009/$R_c$0.006/$I_c0.007$      &80/50/45       & 67          & WHOT   &  Light curve \\
2023 Dec. 5   &10284.92   &10284.99           & V0.006/$R_c$0.006/$I_c0.007$      &80/50/45       & 33          & XL   & Minimum \\
\hline
\end{tabular}}
\end{table}
\end{center}

\begin{figure*}
\begin{minipage}[t]{0.49\textwidth}
\includegraphics[width=\textwidth, angle=0]{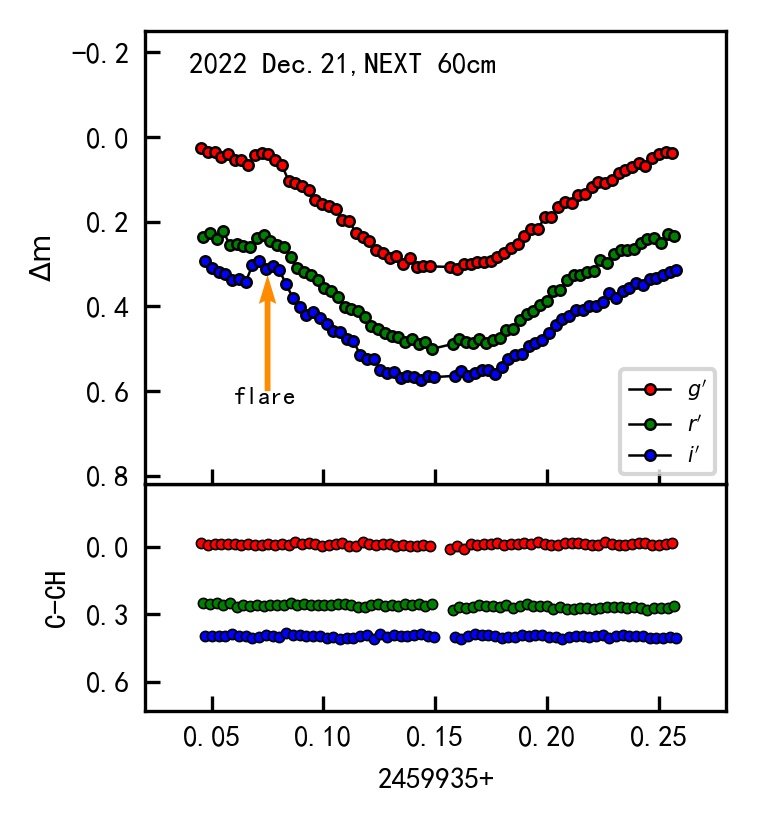}
\end{minipage}
\begin{minipage}[t]{0.49\textwidth}
\includegraphics[width=\textwidth, angle=0]{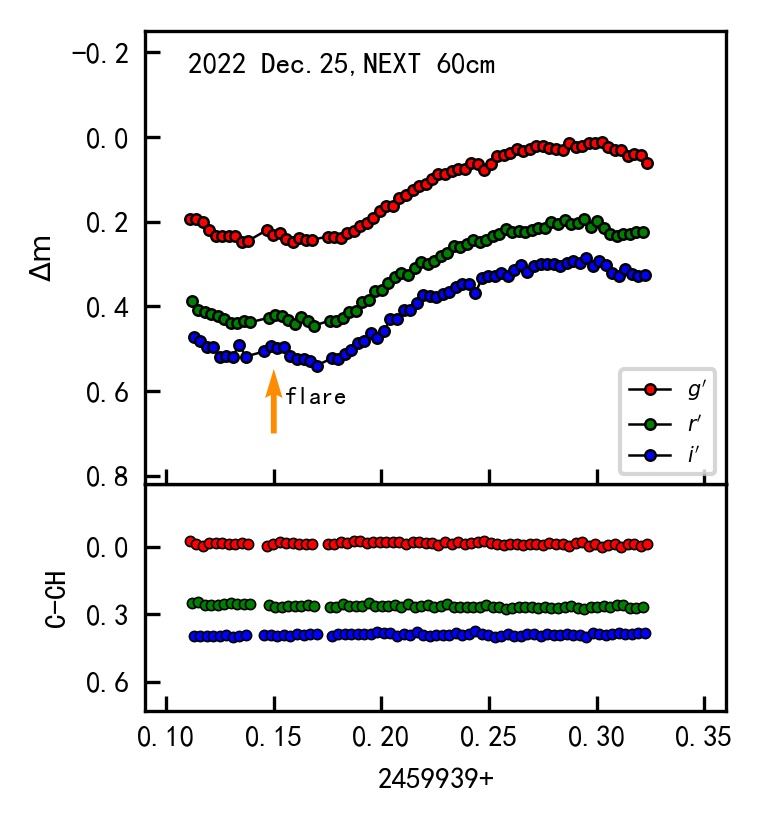}\hspace{-5cm}
\end{minipage}
\caption{Light curves with flares of HAT 307-0007476 on 2022 December 21 (top left), and 25 (top right). The yellow arrows indicate the flare events. \label{fig:flare}}
\end{figure*}

All the CCD images were reduced by IRAF\footnote{IRAF (\url{http://iraf.noao.edu/}) is distributed by the National Optical Astronomy Observatories, which are operated by the Association of
Universities for Research in Astronomy, Inc., under a cooperative agreement with the National Science Foundation.}. The data reduction includes bias and flat corrections, aperture photometry for the target, the comparison and check stars. The differential photometry method was adopted to obtain the light curves. The target's right ascension is $03^h23^m52.51^s$ and declination is $ +18^{\circ}12^{\prime}32.1^{\prime\prime}$, and two stars with similar brightness around the target were selected as the comparison star and the check star, respectively. The same comparison star and check star were selected for image processing for the different telescopes. The comparison star is 2MASS J032346.88+180720.6, and the check star is 2MASS J032347.79+181623.6. The magnitude difference between the comparison star and the check star is constant. We finally obtained two complete sets of light curves covering the entire period: a set of light curve in 2022 by NEXT and a set of light curve in 2023 by WHOT. We detected two significant humps in the set of light curves in 2022 by NEXT. The first hump was detected on the night of December 21, 2022, the light curve of this night is displayed in the left panel of Figure \ref{fig:flare}. The second hump was detected on the night of December 25, 2022, the light curve of this night is displayed in the right panel of Figure \ref{fig:flare}. The top panel of these two figures shows the multiple color light curves of our target, and the bottom panel shows the magnitude difference between the comparison and check stars. As shown in this figure, the magnitude difference between the comparison star and the check star is constant. Therefore, we inferred that the two humps come from the luminosity variation of the binary system. Both the light curves have a rapid rise followed by a slow decay. The definition of flare from the literature is as follows: the light variation must last a few minutes (at least 3 points) with a peak amplitude of not less than 0.03 mg \citep{2015ApJ...814...35C,2016Ap&SS.361..302P,2017MNRAS.466.2542S}.  In addition, the average errors for the observed light curve on December 21 and 25, 2022 are both smaller than 0.007 mag, the peak amplitudes of the two humps are above at least four times sigma levels above the observational errors. Therefore, we infer that the two humps in the light curves are two flare events from this target.

\subsection{TESS Photometry}
\begin{figure}
\centering
\includegraphics[width=0.49\textwidth]{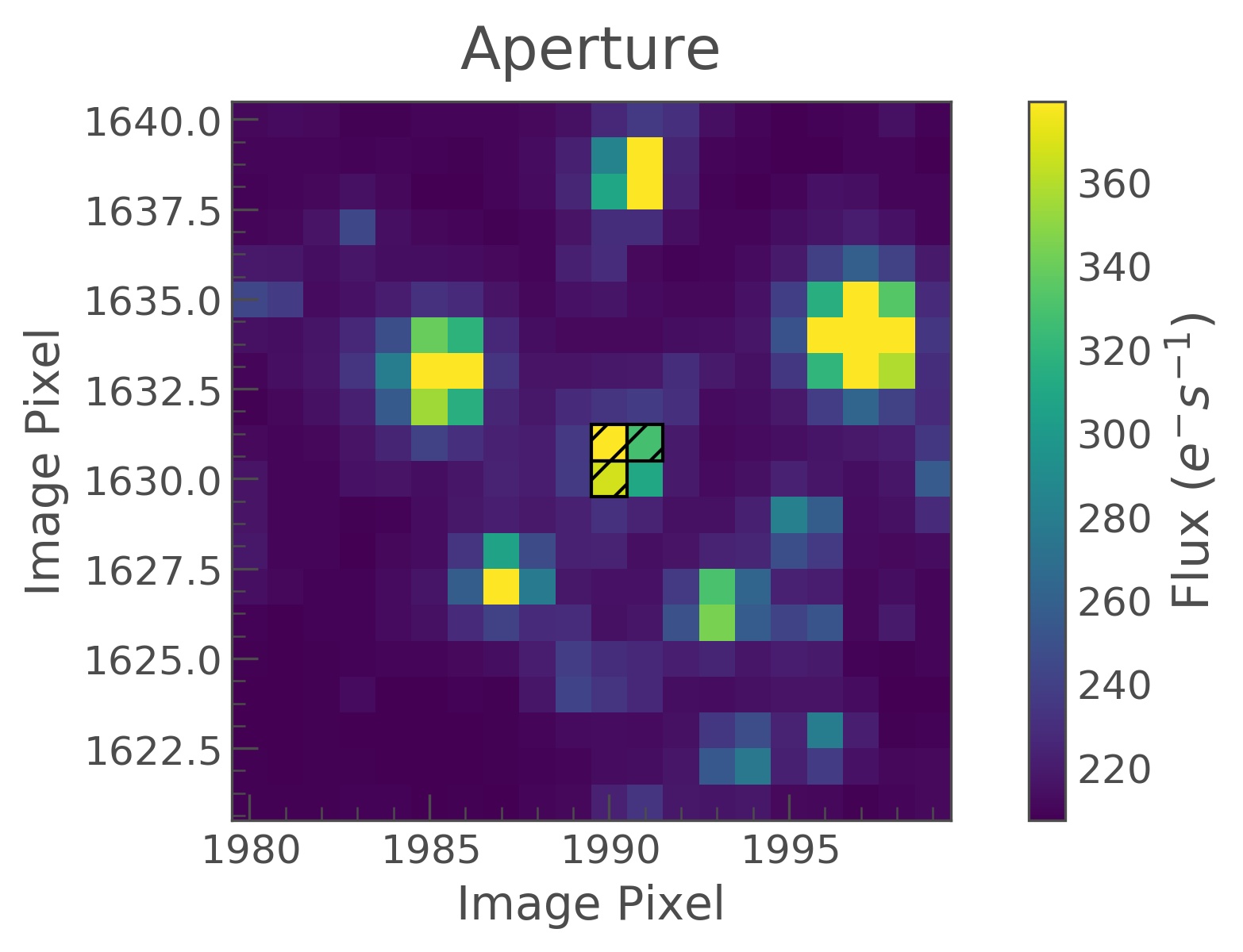}
\includegraphics[width=0.49\textwidth]{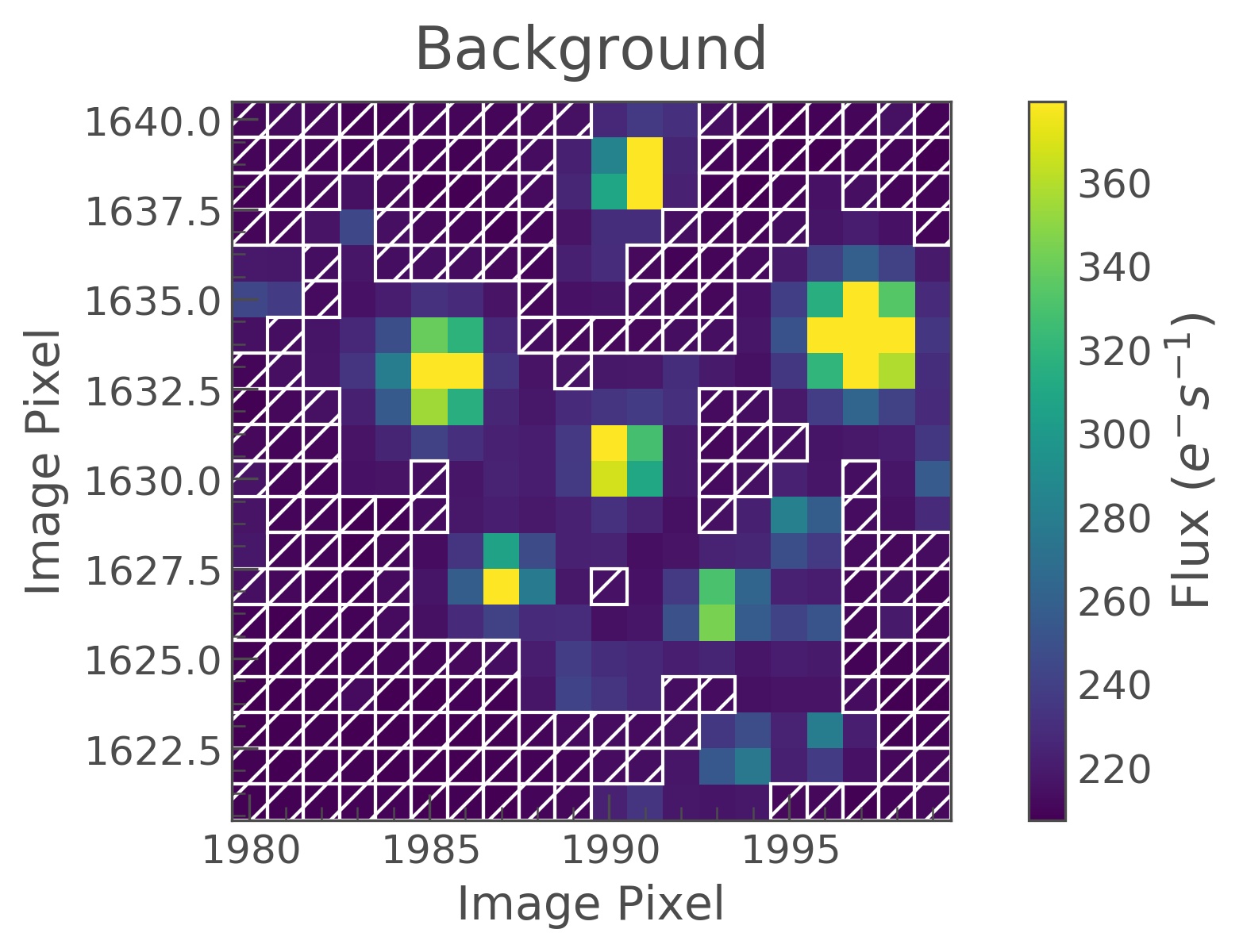}
\caption{The left panel shows the photometric aperture pixels (the rectangular box) selected using lightkurve, and the right panel shows the selected background skylight pixels (the rectangular box). \label{fig:FFI}}
\end{figure}

The Transiting Exoplanet Survey Satellite (TESS) is a space all-sky survey telescope, focusing on stars brighter than approximately 16 magnitude in the TESS band \citep{2015JATIS...1a4003R, 2018AJ....156..132O}. TESS observed HAT 307-0007476 in sectors 42, 43, and 44. We retrieved 20 × 20 pixel cut-outs of TESS Full Frame Image (FFI) for HAT 307-0007476 using its coordinates. This was accomplished via the Lightkurve package, which accesses data available through the Mikulski Archive for Space Telescopes (MAST) \citep{2019ascl.soft05007B}. We selected an appropriate threshold to generate a 2-pixel aperture, as illustrated in the top left panel of Figure \ref{fig:FFI}, where the aperture is indicated by the rectangular box, these three pixels are on either side of each other, so the equivalent aperture is 2-pixel, and a 200-pixel background was selected as shown in the top right panel of Figure \ref{fig:FFI}, where the background is delineated by the rectangular box. These selections were used to extract light curves, incorporating the background subtraction process. Finally, the light curve was de-trended by applying a polynomial fit, and then we normalized the flux. The light curves obtained are in a 10-minute cadence. 

\begin{figure}
\centering
\includegraphics[width=1\textwidth]{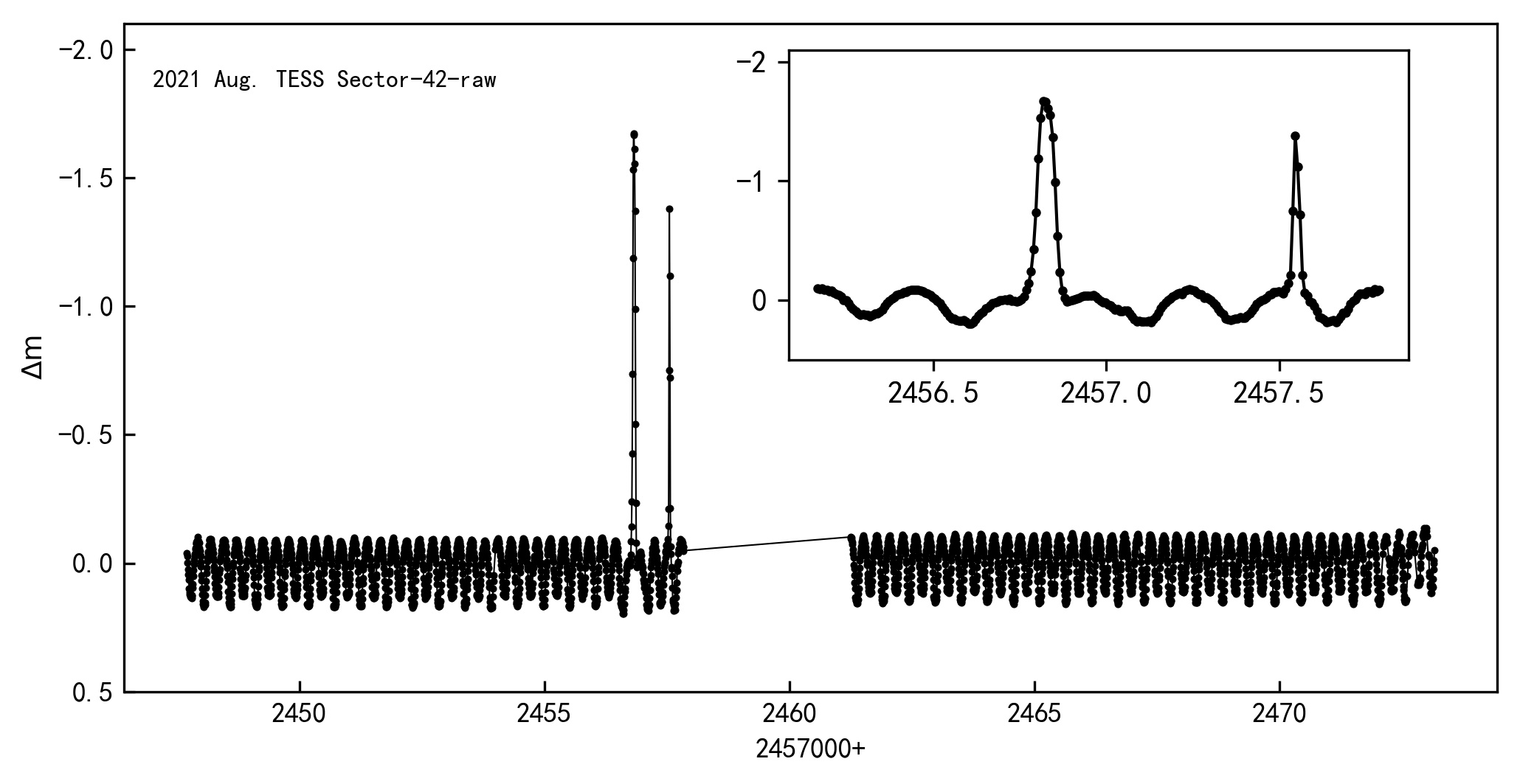}
\caption{The raw complete light curves extracted from the FFI of Sectors 42 and the zoomed-in view of two fake flare events.\label{fig:lightcurve of TESS 42}}
\end{figure}

\begin{figure}
\centering
\includegraphics[width=1\textwidth]{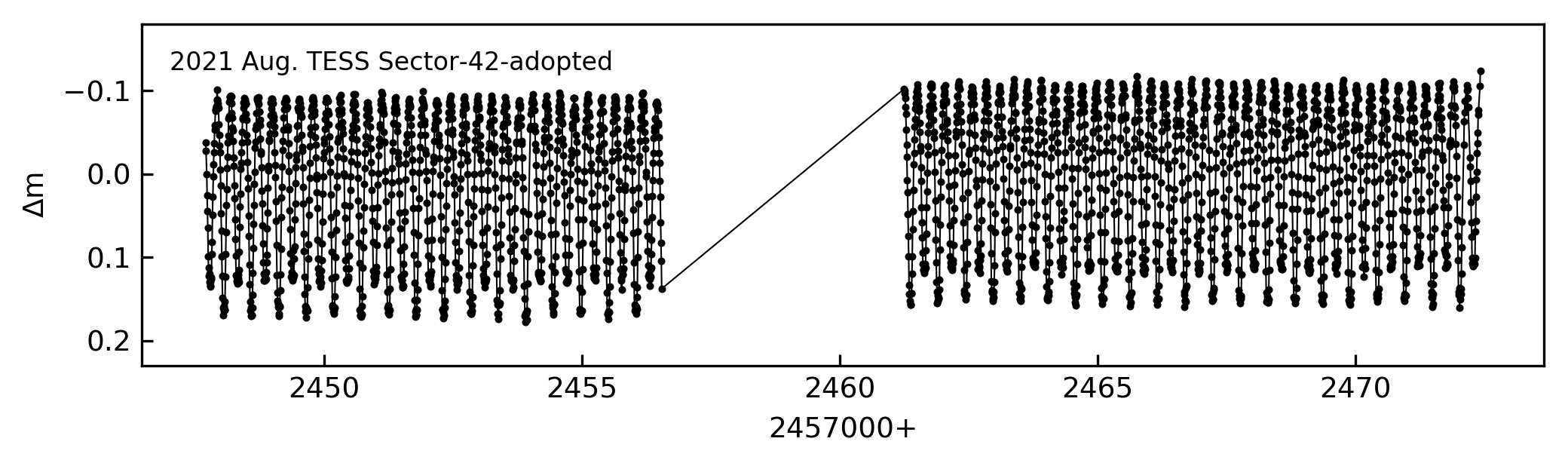}
\includegraphics[width=1\textwidth]{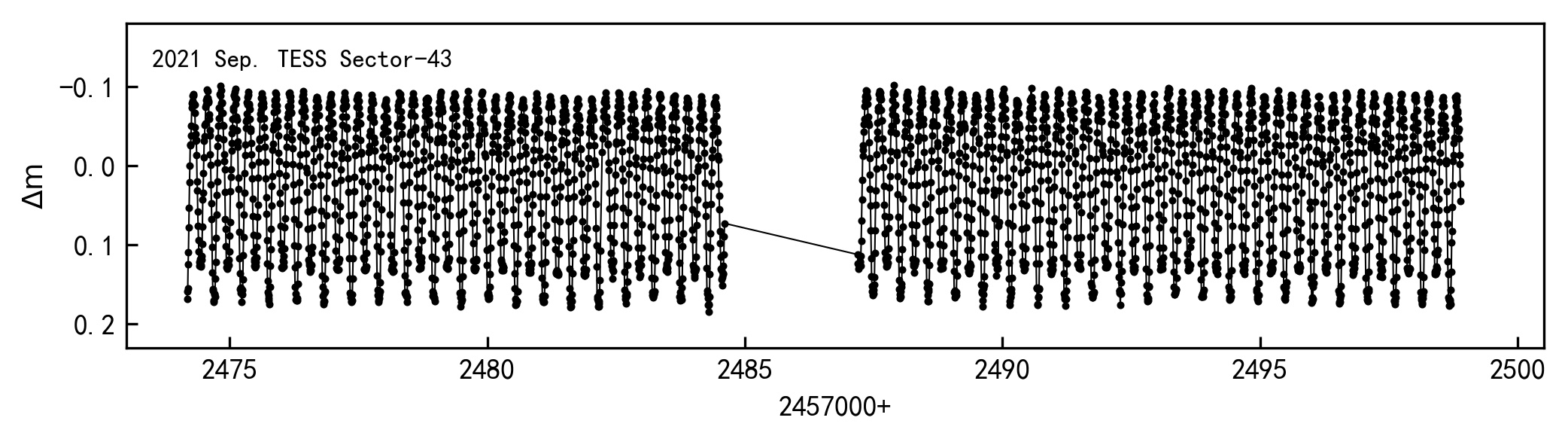}
\includegraphics[width=1\textwidth]{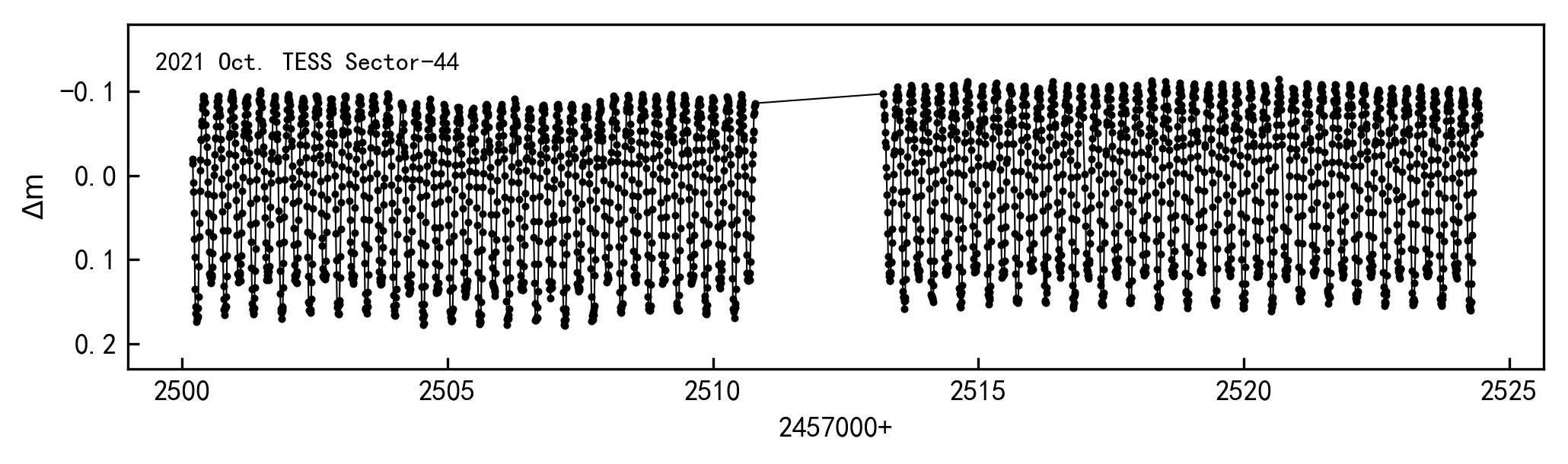}
\caption{The final adopted light curves extracted from the FFI of Sector-42, 43 and 44.\label{fig:lightcurve of TESS}}
\end{figure}

In the raw light curve extracted from the FFI of Sector-42 (see Figure \ref{fig:lightcurve of TESS 42}), there are two quick and huge brightening events, there is a zoomed-in view of this part of the light curve in the upper right. Initially, we suspected that the sudden brightening was due to a flare of our target, since the light curve is characterized by a rapid rise and a slow decay. However, after careful analysis, we removed the two fake flare signals. The first event around 2459456.8 (BJD) is not only the transit of asteroid 219 Thusnelda (also known as A880 SA, and it is one magnitude brighter than our target) through the aperture according to the online URL\footnote{\url{https://www.minorplanetcenter.net/iau/mpc.html}}, but was also disturbed by stray light according to the data quality flags of the FITS files. The second event in the next day is caused by stray light or some instrumental reflection according to the data quality flags of the FITS files. We removed these points with non-zero quality flags in the FITS files. The final adopted light curve of Sector-42 is shown in the first panel of Figure \ref{fig:lightcurve of TESS} and the light curves of Sector-43 and 44 are shown in the second and third panel of Figure \ref{fig:lightcurve of TESS}. Since TESS pixels are large ($21^{\prime\prime}$), crowdings may be an issue. There are only two very faint stars (one fainter than 20 mag and the other fainter than 18 mag) within $42^{\prime\prime}$ from the target in the Gaia-DR3 catalog, equivalent to 2 TESS pixels. Our target is 100-1000 times more luminous than these two faint stars. Therefore, the light curves appear representative of the HAT 307-0007476 itself.

\subsection{LAMOST Spectroscopy}
HAT 307-0007476 was observed with low-resolution spectroscopy in four epochs by the Large Sky Area Multi-Object Fiber Spectroscopic Telescope (LAMOST) \citep{2012RAA....12.1197C,2015RAA....15.1095L}. LAMOST is a 4-m reflecting Schmidt telescope featuring a field of view of $5^\circ$. The focal plane is equipped with 4000 fibers, enabling a high spectral acquisition rate. The low-resolution spectroscopic survey achieves a spectral resolution of $R\sim$ 1800, covering a wavelength range from approximately 370 to 900 nm. The details of the LAMOST spectra are summarized in Table \ref{table:LAMOST}.

\begin{table}
\caption{The Information of LAMOST Spectra} \label{table:LAMOST}
\begin{tabular}{ccccccccc}
\hline\hline
    UT Date     &Phase & Sp &$SNR_g$  & T$_{eff}$ & log g  & [Fe/H] & RV  &H$_\alpha$ \\
  && & & (K) & (dex) & (dex)&(km/s) & ({\AA})\\
\hline
2015-12-20 &0.69106     & F0 &20 & 6842.87($\pm$68) & 4.140($\pm$0.110) & -0.164($\pm$0.072) & -7.01($\pm$5.65)& 0.373($\pm$0.016)  \\
2015-12-27  &0.80200    & F0 &24 & 6821.39($\pm$64) & 4.262($\pm$0.100) & -0.220($\pm$0.067) & -6.89($\pm$6.10)& 0.258($\pm$0.031)  \\
2016-01-12  &0.72715    & A9 &30 & 6820.34($\pm$155)& 4.185($\pm$0.254) & -0.363($\pm$0.150) &-20.15($\pm$7.29)& 0.296($\pm$0.025)  \\
2016-11-04  &0.44793    & F0 &56 & 6894.56($\pm$35) & 4.119($\pm$0.050) & -0.205($\pm$0.030) & 31.91($\pm$4.69)& 0.303($\pm$0.005)  \\

\hline
\end{tabular}
\end{table}

\section{Investigations} \label{sec:investigation}
\subsection{Photometric investigation}
The light curves of HAT 307-0007476 have a flat-bottom minimum, so it is a totally eclipsing system. Based on the statistical analysis, \cite{2003CoSka..33...38P} proposed the mass ratio of the photometric investigation and the spectroscopic investigation are almost equal when light curves have flat-bottom minima. This means that the mass ratio obtained from photometric data alone is reliable when spectroscopic data are unavailable. \cite{2005Ap&SS.296..221T} verified the results of this statistical analysis based on the numerical model. \cite{2017MNRAS.466.1118Z} proposed the symmetrical light curve and the sharp bottom of the q-search curve as two indications for obtaining a reliable mass ratio. In conclusion, when the radial velocity curve is unavailable, the physical parameters of the contact binaries are reliable only in the totally eclipsing systems \citep{2021AJ....162...13L}.

We used the q-search method to get the mass ratio, $q=m_2 / m_1$. The q-search method is a widely applied approach to estimate contact binary mass ratios without the radial velocity curves \citep{1997A&AS..124..291N,2003A&A...403..675O,2019PASP..131e4203C,2020ApJS..247...50S}. We started with light curves without flare activities of NEXT, WHOT, and TESS in seven bands simultaneously to improve the reliability of the mass ratio calculation. In this case, the data of TESS is binned to 1000 points by averaging, which, in addition to save computational time, the number of points in the light curve of the binned TESS is consistent with the number of points in the light curve of NEXT and WHOT, and the constraints on the numerical model are balanced. All light curves are phase folded using the following linear ephemeris,

\begin{eqnarray}\label{E0}
%\begin{aligned}
Min. I = &T_0+0.^d5329326\times E,
%\end{aligned}
\end{eqnarray}
where the $T_0$ is the initial primary minimum that is obtained using the K-W method \citep{1956BAN....12..327K} based on NEXT, WHOT, and TESS photometric observations data, listed in Table \ref{tab:minima}, the orbital period 0.5329326 is from the ASAS-SN database, and $E$ is the cycle number.

\begin{figure*}
\centering
\includegraphics[width=0.8\textwidth]{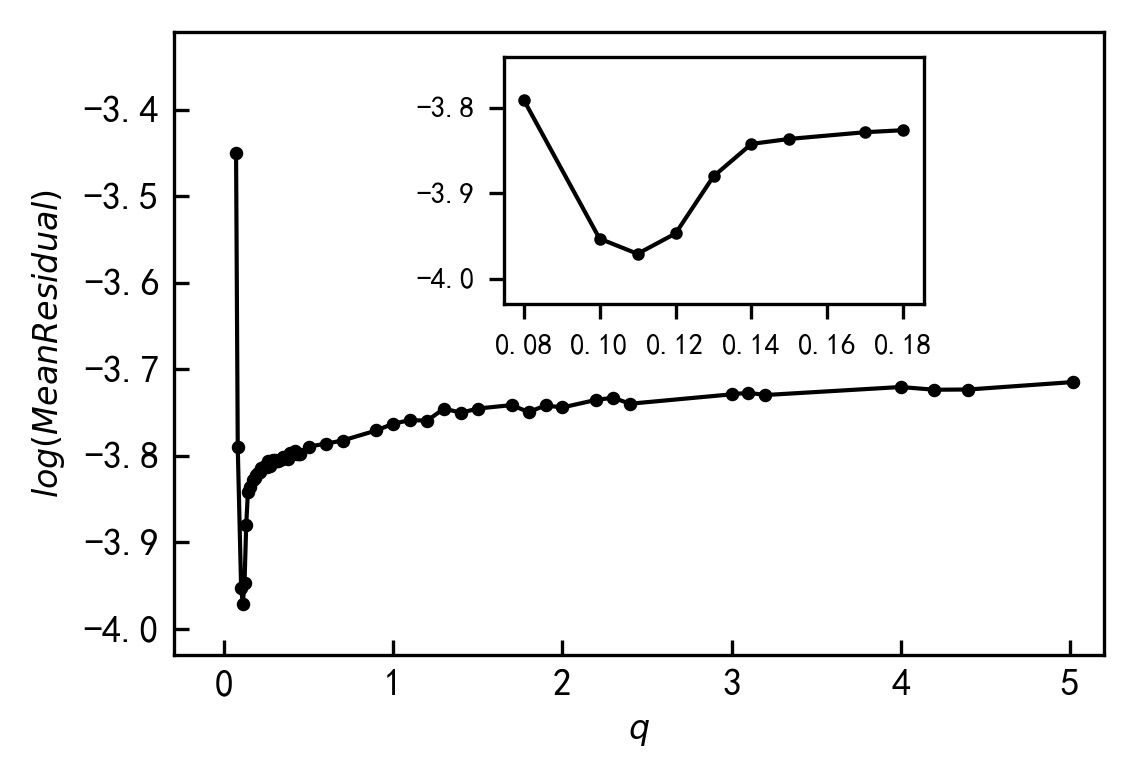}
\caption{The relationship between mean residual and the mass ratio q. The small diagram in the figure is a zoomed-in view near the optimal solution..}
\label{qsearch}
\end{figure*}

\begin{figure}
\centering
\includegraphics[width=0.8\textwidth]{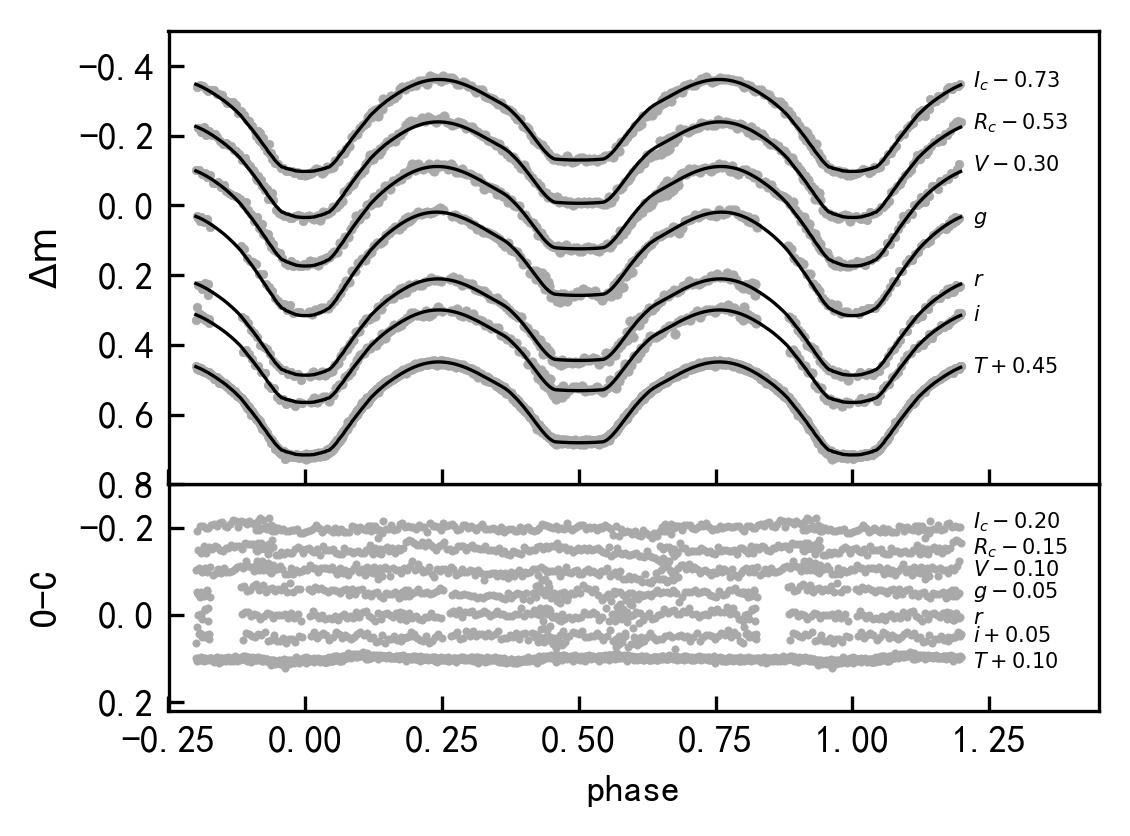}
\caption{The theoretical light curves (continuous lines) with simultaneous solution and observed light curves from WHOT, NEXT, and TESS.}
\label{fig:solution}
\end{figure}

The program we used for analysis was W-D code 2015 version \citep{1971ApJ...166..605W,1979ApJ...234.1054W,1990ApJ...356..613W}.
Our criterion for determining the optimal solution of $q$ was that the mean residual at that q-value was minimum; the mean residual is the square root of the mean-weighted sum of squares of the residuals according to the \cite{1971ApJ...166..605W, 1979ApJ...234.1054W, 1990ApJ...356..613W}. “Mean residual” has been widely applied in many papers (e.g. \citealt{2016AJ....152..129C, 2019PASP..131e4203C, 2020ApJS..247...50S}). Searching for different $q$ values, from 0.05 to 5. The step is set to 0.01 when $q$ is less than 0.5, and 0.1 when $q$ is greater than 0.5. During this process, we adopted “Mode 3,” appropriate for over-contact binaries, with both component stars filling their Roche lobes. At first, we take the average of four temperature values listed in Table \ref{table:LAMOST} obtained by LAMOST as the primary effective temperature. Based on the primary effective temperature, we fixed the gravity darkening and bolometric albedo coefficients at $g_1=g_2=0.32$ and $A_1=A_2=0.5$ \citep{1924MNRAS..84..665V}, and the limb-darkening coefficients were interpolated from \cite{1993AJ....106.2096V}'s table due to a logarithmic law (ld=-2) following \cite{2020Obs...140..247S,2023A&A...671A.139P,2024MNRAS.527.4076K}. The numerical grid size was set to N$_1$=N$_2$=60 and N$_{1L}$=N$_{2L}$=50 following \cite{2020Obs...140..247S,2023A&A...671A.139P}. The parameters fitted by W-D code are orbital inclination in degrees, i; average surface temperature of secondary, $T_2$; surface potential, $\Omega_1=\Omega_2$; the relative monochromatic luminosity of primary, $L_1$. The result of the q-search is shown in Figure \ref{qsearch}. It is clear that the minimum mean residual corresponds to the optimal solution of $q=0.11$. After obtaining a reliable mass ratio, the mass ratio is set as an adjustable parameter, and the values of the other physical parameters are obtained by calculation. This result is listed in the second column of Table \ref{table:all solutions}. The theoretical light curves with simultaneous solution and observed light curves are presented in Figure \ref{fig:solution}.
 
After obtaining the simultaneous solution, the differences in the light curves can lead to different values of other physical parameters. Therefore, we subsequently separated NEXT and WHOT from the TESS data and independently determined the other physical parameters.

\begin{figure*}
\begin{minipage}[t]{0.49\textwidth}
\includegraphics[width=\textwidth, angle=0]{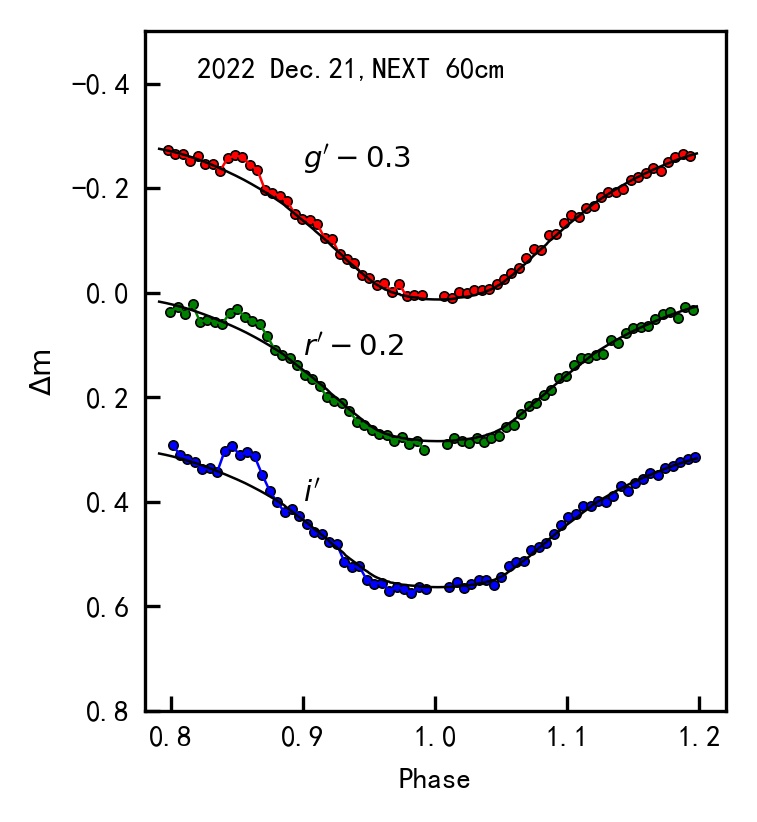}
\end{minipage}
\begin{minipage}[t]{0.49\textwidth}
\includegraphics[width=\textwidth, angle=0]{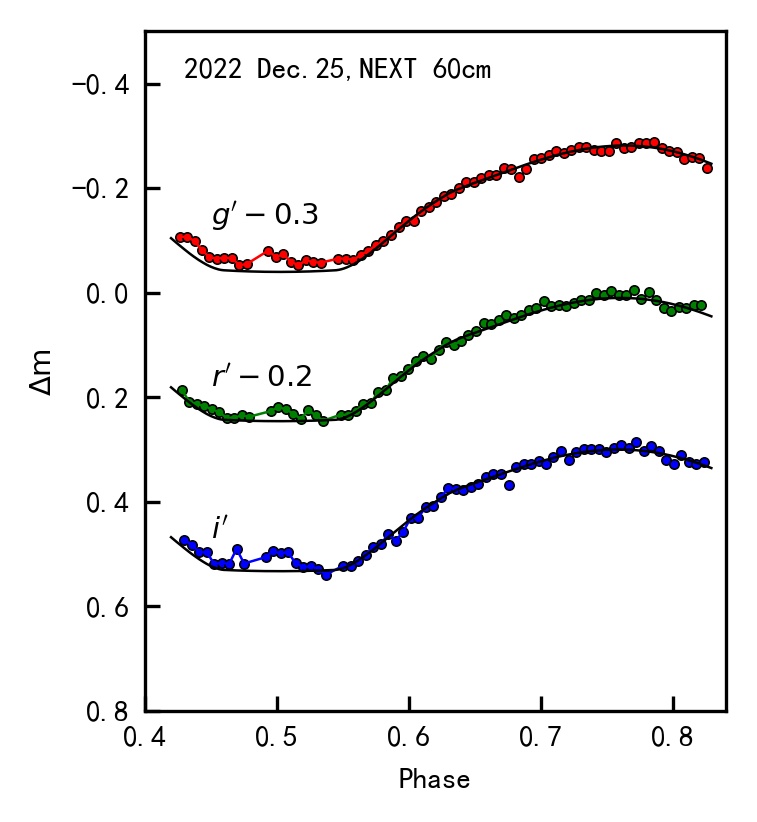}\hspace{-5cm}
\end{minipage}

%\begin{minipage}[t]{1\textwidth}
%\includegraphics[width=\textwidth, angle=0]{tess_s42_flares.jpg}
%\end{minipage}
\caption{Theoretical light curves and observed curves with flare activities for HAT 307-0007476. The solid line indicates the best-fit light curve. \label{fig:flares}}
\end{figure*}

\begin{figure*}
\centering
\begin{minipage}[t]{0.325\textwidth}
\centering
\includegraphics[width=\textwidth, angle=0]{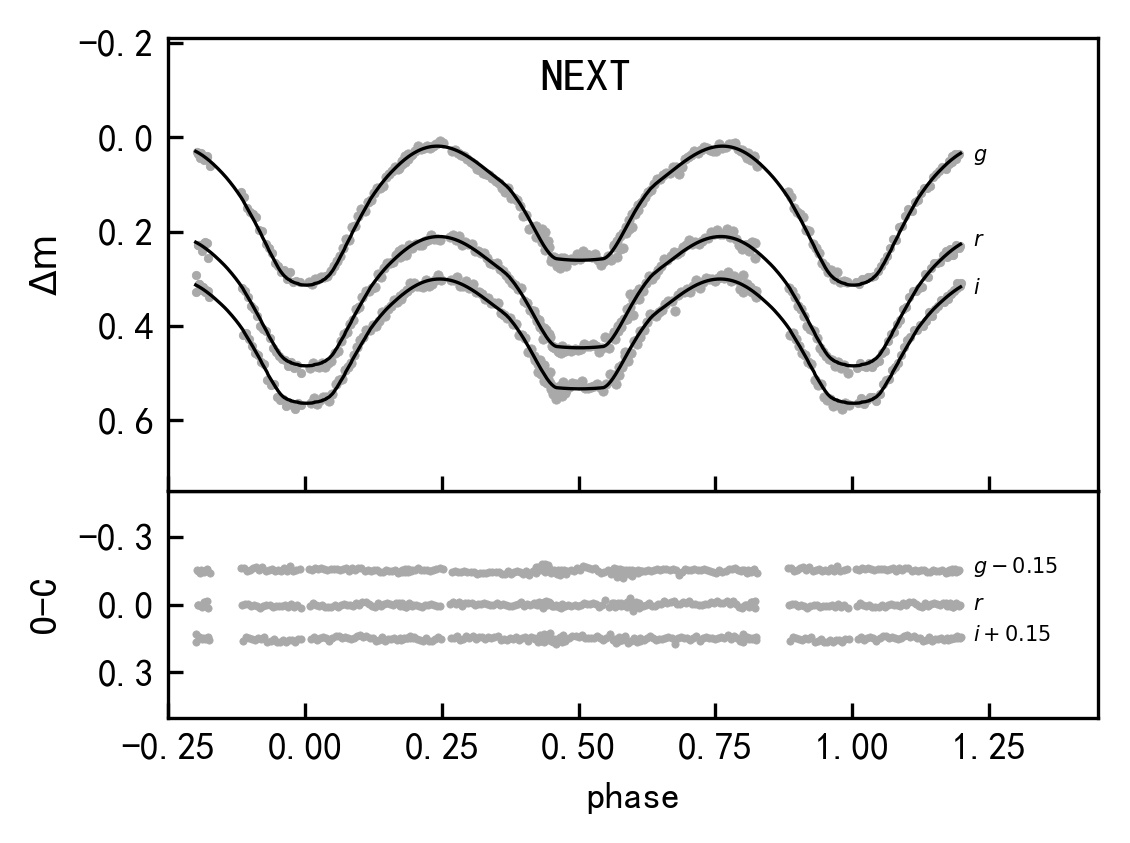}
\end{minipage}
\begin{minipage}[t]{0.325\textwidth}
\centering
\includegraphics[width=\textwidth, angle=0]{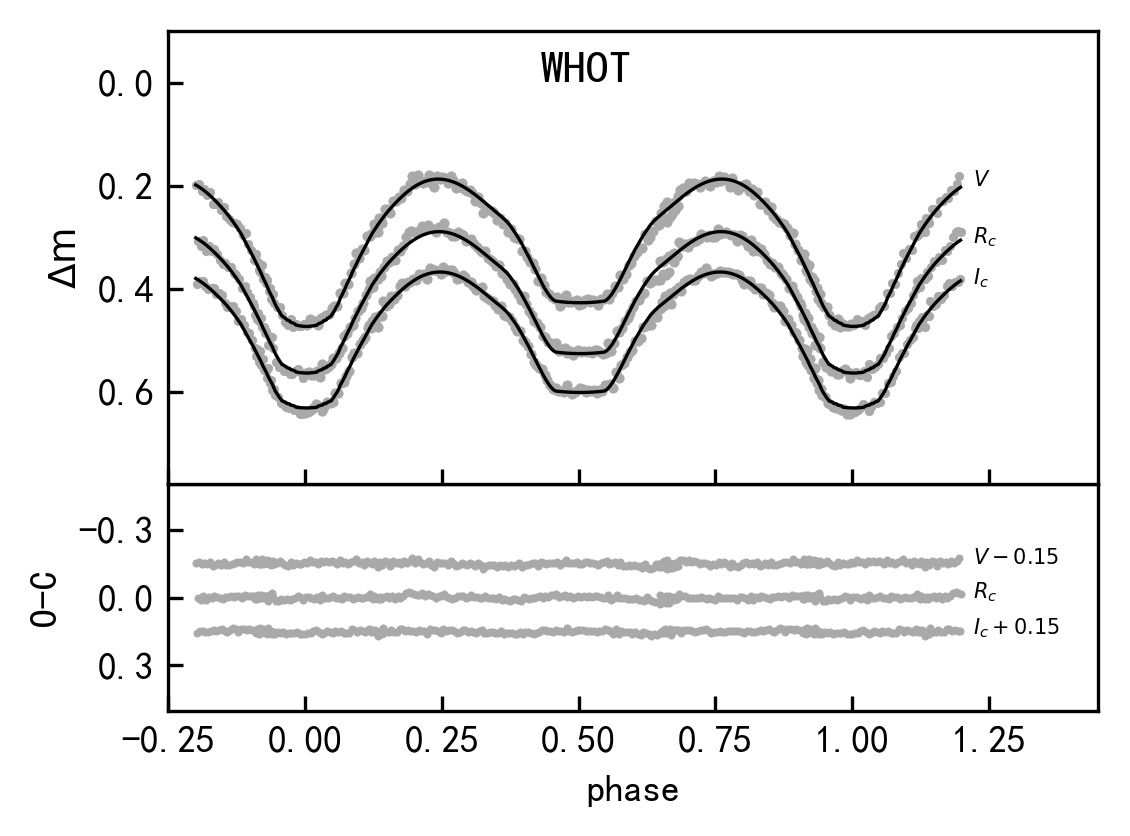}
\end{minipage}
\begin{minipage}[t]{0.325\textwidth}
\centering
\includegraphics[width=\textwidth, angle=0]{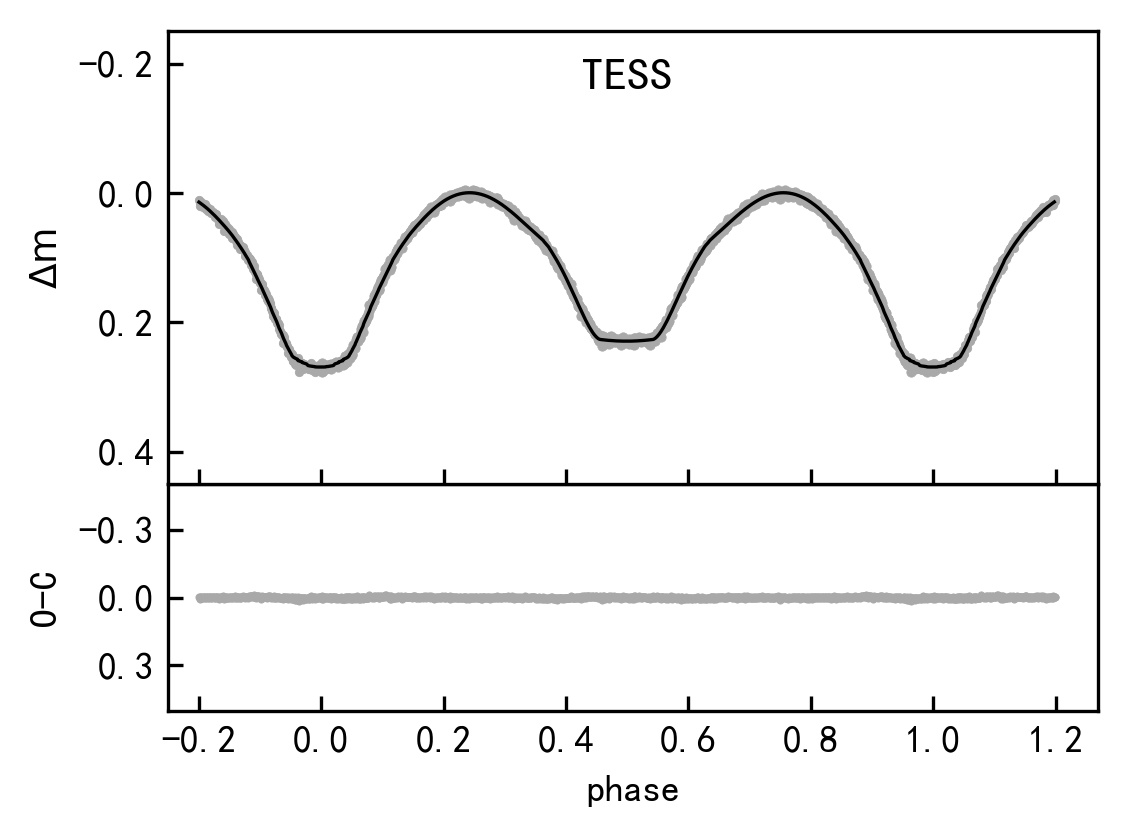}
\end{minipage}
\begin{minipage}[t]{0.325\textwidth}
\centering
\includegraphics[width=\textwidth, angle=0]{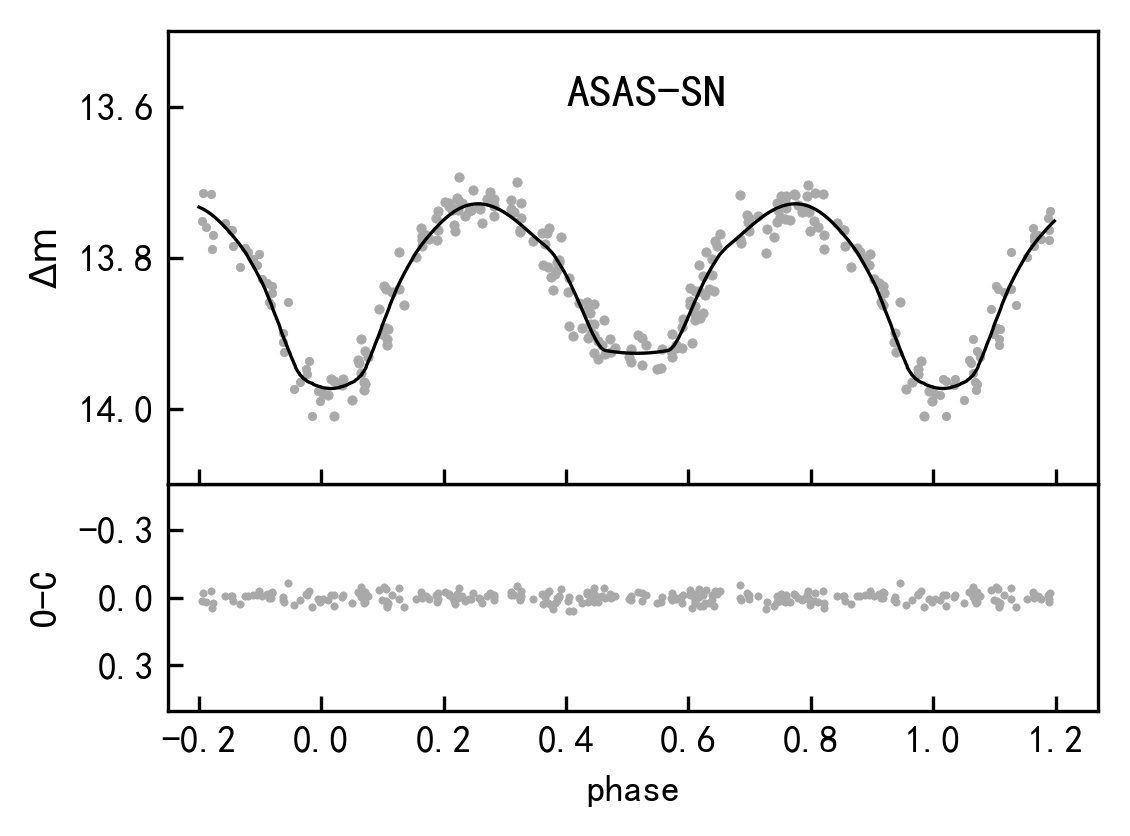}
\end{minipage}
\begin{minipage}[t]{0.325\textwidth}
\centering
\includegraphics[width=\textwidth, angle=0]{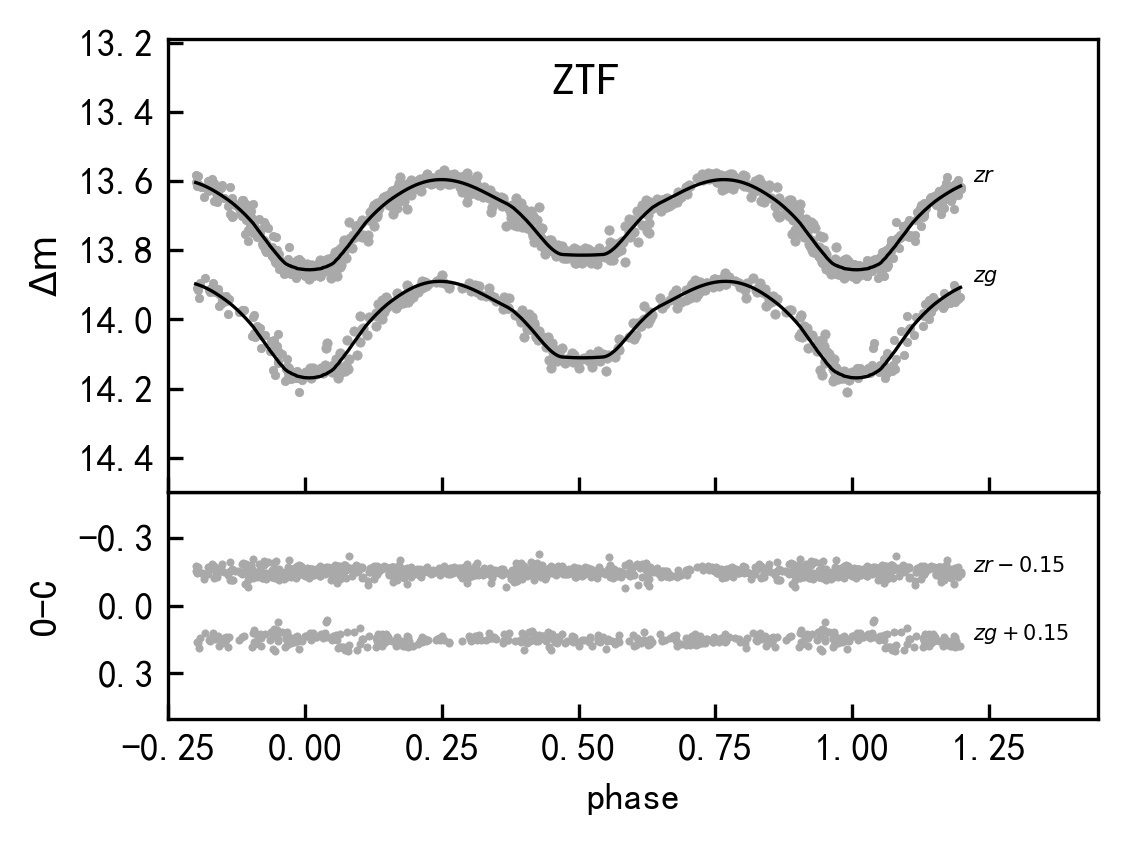}
\end{minipage}
\begin{minipage}[t]{0.325\textwidth}
\centering
\includegraphics[width=\textwidth, angle=0]{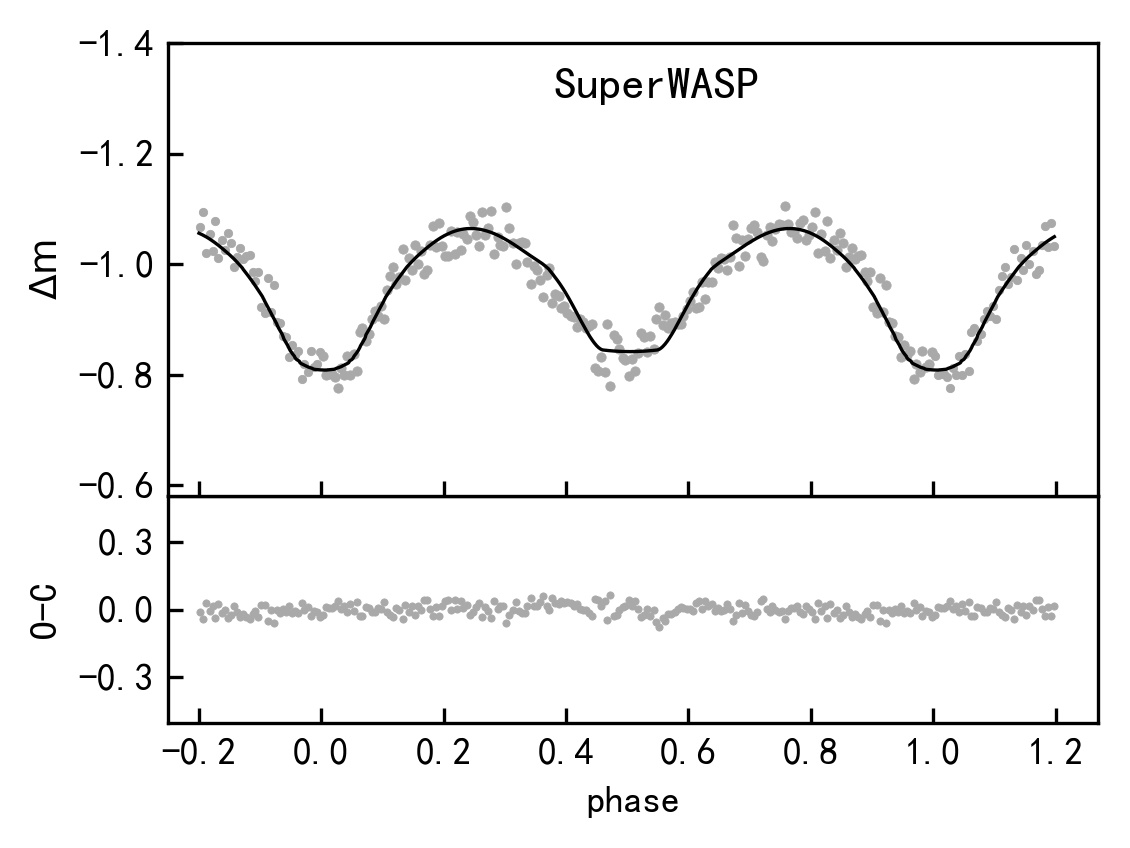}
\end{minipage}
\begin{minipage}[t]{0.33\textwidth}
\centering
\includegraphics[width=\textwidth, angle=0]{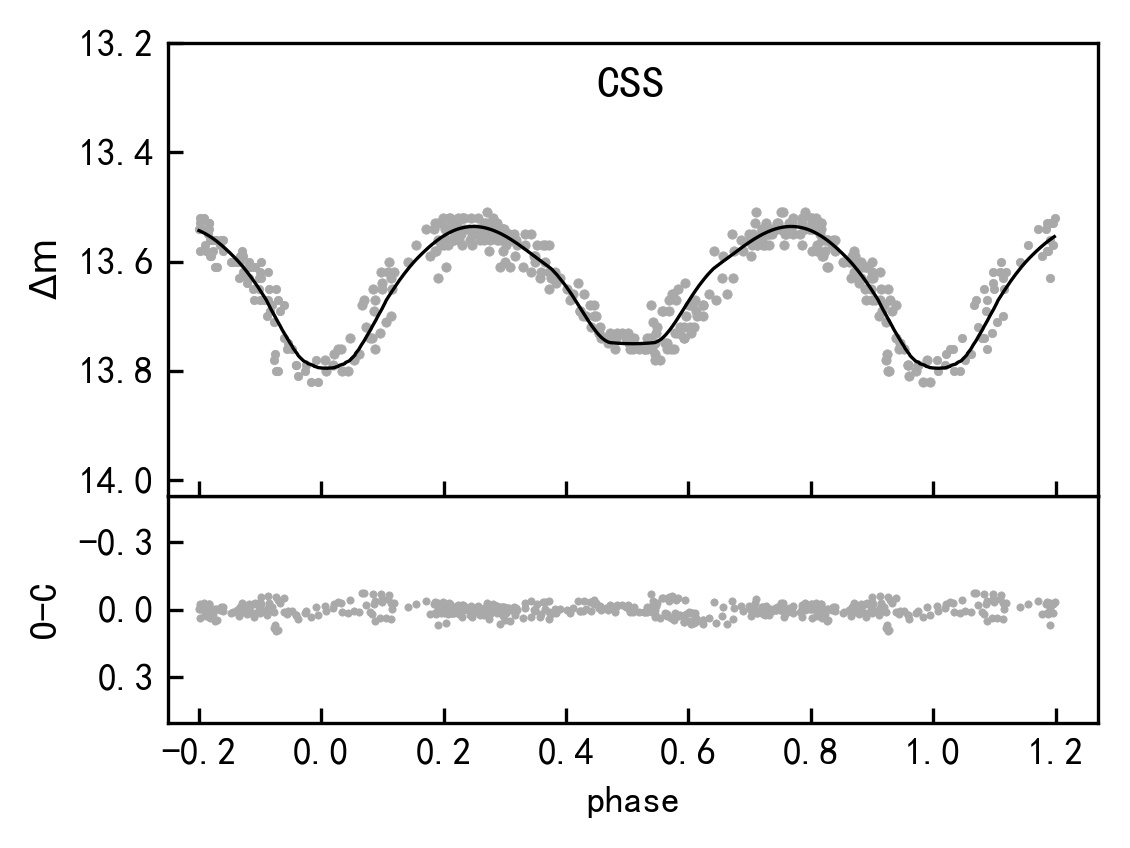}
\end{minipage}
\caption{Theoretical light curves and observed light curves of HAT 307-0007476. \label{fig:LCs}}
\end{figure*}

Then, we tried to determine the optimal value of the other physical parameters independently using the NEXT, WHOT, and TESS data with $q=0.11$ as the initial value to obtain photometric solutions for each physical parameter. The mass ratio was set as an adjustable parameter. After several iterations, the convergent solution was obtained. The solutions of these three sets of light curves are labeled solution-NEXT, solution-TESS, and solution-WHOT, respectively. All the parameters are shown in Table \ref{table:all solutions}, and the physical parameters derived from TESS data is adopted as the final results since the sensitivity of TESS data should be significantly superior compared to other light curves.

Then, the photometric solutions of the light curves of HAT 307-0007476 with flare activities were calculated. We used solution-NEXT to analyze the light curves with flare activities for December 21 and 25, 2022. Figure \ref{fig:flares} presents the best simulation results; the observed curve is represented by dots and the theoretical curve is shown as a solid black line.

At last, we also analyzed photometric data from multiple sky surveys, including the ASAS-SN \citep{2014ApJ...788...48S,2017PASP..129j4502K,2018MNRAS.477.3145J}, the Zwicky Transient Facility (ZTF; \citealt{2019PASP..131a8002B, 2019PASP..131a8003M}), the Wide Angle Search for Planets (SuperWASP; \citealt{2006PASP..118.1407P, 2010A&A...520L..10B}), and the CSS \citep{2009ApJ...696..870D} with the simultaneous solution as the initial value. This approach offers a more comprehensive understanding of the target by comparing data from different sources. Because the SuperWASP light curve is relatively scattered, we obtained the rectified light curve by averaging the 4000+ points and averaged them to 200 points. Best-fitting theoretical light curves (continuous lines) are also shown in Figure \ref{fig:LCs}. The solutions based on photometric data from different sky surveys are listed in Table \ref{table:all solutions}, too. All of these solutions are fairly consistent.

\begin{center}
\begin{table}\centering
\footnotesize
\caption{The photometric solutions of HAT 307-0007476}
\label{table:all solutions}
\setlength{\tabcolsep}{1mm}{
\begin{tabular}{lcccccccc}
\hline\hline
Parameters & Simultaneously & NEXT & WHOT & TESS & ASAS-SN & ZTF & SuperWASP & CSS\\
\\\hline
	
$T_1(K)$                      &6845 &6845 &6845 &6845 &6845 &6845 &6845 &6845    \\
$q(M_2/M_1) $                 &0.109$\pm0.001$  &0.108$\pm0.002$ &0.110$\pm0.005$  &0.114$\pm0.001$ &0.096$\pm0.004$ &0.105$\pm0.002$ &0.101$\pm0.005$ &0.096$\pm0.004$\\
$T_2(K)$                      &6547$\pm8$     &6597$\pm15$    &6583$\pm16$   &6474$\pm6$ &6442$\pm68$ &6467$\pm25$ &6639$\pm78$ &6495$\pm63$\\
$i(deg)$                      &75.5$\pm0.2$     &75.7$\pm0.4$    &75.9$\pm0.4$   &76.3$\pm0.1$ &78.2$\pm2.3$ &74.1$\pm0.6$ &75.8$\pm2.1$ &74.3$\pm2.9$\\
$\Omega_1=\Omega_2$           &1.954$\pm0.003$  &1.952$\pm0.006$ &1.958$\pm0.006$ &1.973$\pm0.002$ &1.941$\pm0.018$ &1.947$\pm0.009$ &1.948 $\pm0.022$ &1.921$\pm0.020$ \\
$L_{1}/(L_{1}+L_{2})(g^{'})$      &0.895$\pm0.001$  &0.892$\pm0.001$  & $-$ & $-$  & $-$ & 0.905$\pm0.003$ & $-$ & $-$         \\
$L_{1}/(L_{1}+L_{2})(r^{'})$      &0.889$\pm0.001$  &0.887$\pm0.001$  & $-$ & $-$  & $-$ &0.898$\pm0.002$ & $-$ & $-$       \\
$L_{1}/(L_{1}+L_{2})(i^{'})$      &0.886$\pm0.001$ &0.885$\pm0.001$ & $-$ & $-$  & $-$ & $-$  & $-$ & $-$         \\
$L_{1}/(L_{1}+L_{2})(V)$          &0.892$\pm0.001$  & $-$  &0.889$\pm0.001$ & $-$ &0.913$\pm0.007$ & $-$ & $-$ &0.905$\pm0.007$        \\
$L_{1}/(L_{1}+L_{2})(R_c)$        &0.889$\pm0.001$  & $-$  &0.886$\pm0.001$ & $-$ & $-$ & $-$ & $-$ & $-$         \\
$L_{1}/(L_{1}+L_{2})(I_c)$        &0.886$\pm0.001$  & $-$  &0.883$\pm0.001$ & $-$ & $-$ & $-$ & $-$ & $-$         \\
$L_{1}/(L_{1}+L_{2})(TESS)$       &0.887$\pm0.001$ & $-$  & $-$  &0.889$\pm0.001$  & $-$ & $-$ & $-$ & $-$ \\
$L_{1}/(L_{1}+L_{2})(SuperWASP)$    &$-$ & $-$ & $-$ &$-$ & $-$ & $-$ &0.897$\pm0.006$ & $-$ \\
$r_1$                         & 0.587$\pm0.001$ &0.588$\pm0.002$ & 0.586$\pm0.002$ &0.582$\pm0.001$ &0.586$\pm0.006$ &0.588$\pm0.002$ &0.586$\pm0.008$ &0.586$\pm0.007$ \\
$r_2$                         & 0.228$\pm0.006$ &0.228$\pm0.015$ & 0.228$\pm0.011$ &0.227$\pm0.004$ &0.206$\pm0.039$ &0.223$\pm0.015$ &0.214$\pm0.041$ &0.214$\pm0.048$\\
$f$                           & 46.6$\pm3.6$\%  &45.4$\pm8.6$\%  & 45.3$\pm7.7$\% &37.1 $\pm0.5$\% &9.3 $\pm28.7$\% &39.4$\pm12.5$\% &22.1$\pm33.8$\% &43.4$\pm31.3$\%\\
\hline
\end{tabular}}
\end{table}
\end{center}
\subsection{Spectroscopic investigation}
Contact binaries often exhibit magnetic activity such as spots, flares, or plages, which are related to chromosphere activity. Excess emission from the H$\alpha$ line is one important indicator of chromospheric activity \citep{2018A&A...615A.120P,2020MNRAS.495.1252Z,2021MNRAS.506.4251Z}. We used the spectral subtraction technique to subtract the photospheric spectra from the system spectra to obtain emission \citep{1985ApJ...295..162B}. 

\begin{figure}
\centering
\includegraphics[width=0.8\textwidth]{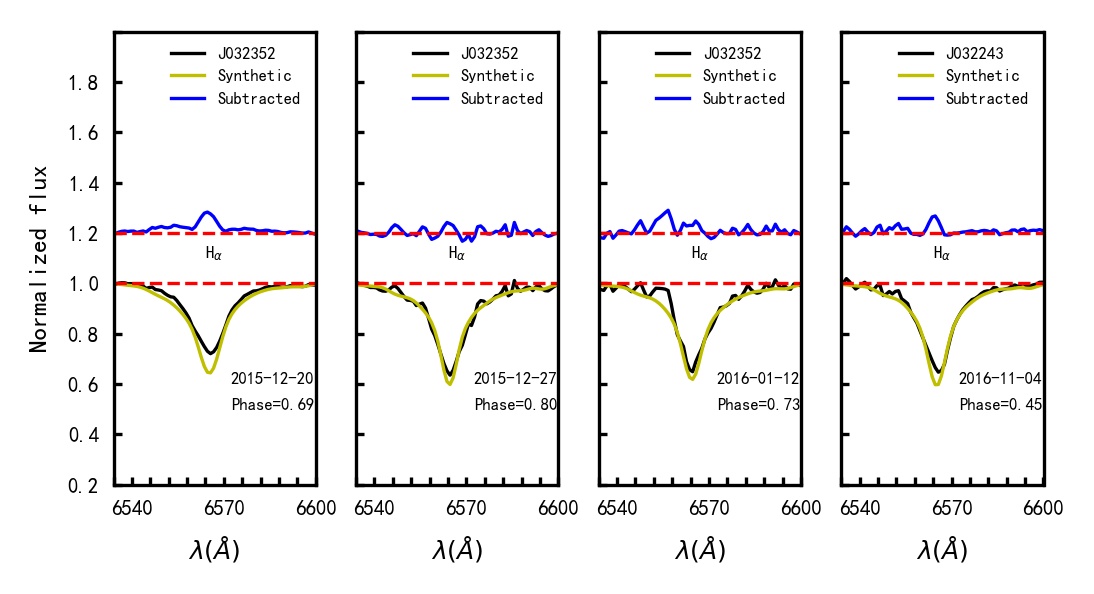}
\caption{The H$_\alpha$ region of low-resolution LAMOST spectra (black line) and synthetic spectra (yellow line) for the HAT 307-0007476 is shown. The continuous blue line in each plot shows the subtracted spectra in the same region. \label{fig:spec}}
\end{figure}

Firstly, we selected the template spectra of the components of this binary system. The template spectrum of the primary component is from inactive star TYC 3128-1088-1, and the template spectrum of the secondary component is from inactive star TYC 1877-1578-1, which are both from the catalog of \cite{2018AJ....156...90H}, the atmospheric parameters of them are shown in Table \ref{table:two templates} . We downloaded the low-resolution LAMOST spectra of the template spectra. From Table \ref{table:two templates, we can see that the} temperature difference between the two inactive stars and the components of binary is less than 200K. Secondly, we normalized the template spectra and the target spectra and removed cosmic rays. Thirdly, we constructed low-resolution synthetic spectra of a binary system. The Fortran code STARMOD developed by \cite{1985ApJ...295..162B} was adopted, and it has been used by many researchers (e.g.\citealt{2000A&AS..146..103M,2010A&A...524A..97L,2022ApJ...927...12P}). 
We synthesized a composite spectrum using the normalized spectra of the two template stars by considering radial velocities, rotational velocities, and the luminosity ratio of the two templates. The initial radial velocities and rotational velocities of them were set as the same of the target binary and its component stars, and the luminosity ratio was used the result determined by the light curve analysis. After 100 iterations, the synthetic spectrum was constructed, and the measured radial velocity, rotation velocity (v$\sin i$), and the flux contribution for the primary are shown in Table \ref{table:two templates}. Finally, we calculated the subtracted spectra. The subtracted spectra between the LAMOST spectra and the synthetic spectra are shown in Figure \ref{fig:spec} (the H$_\alpha$ line region is shown).

Figure \ref{fig:spec} displays the low-resolution LAMOST and synthetic spectra for HAT 307-0007476 and the subtracted spectra in the same region. Clear emission lines can be seen in the first and fourth subplots, which indicates their chromospheric activity, and the spectral emission lines of the two middle subplots are not obvious. The phases of the observations for each spectrum are labeled in the figure. The equivalent widths of excess emissions are listed in Table \ref{table:LAMOST}. 
This result should be considered as a preliminary result due to the limitations of the low-resolution observations, so more high-resolution spectra are needed in the future to help to analyze the emission line sources. 

\begin{table}
\caption{The atmospheric parameters of the two template stars and the derived parameters of the primary component} \label{table:two templates}
\begin{tabular}{cccccccc}
\hline\hline
Star&    UT Date     & Sp &$SNR_g$  & T$_{eff}$ & log g  & [Fe/H] & RV \\
  && & & (K) & (dex) & (dex) & (km/s)\\
\hline
TYC 3128-1088-1  &2015-09-25 & F0 &389 & 6930.68($\pm$8.90) & 4.090($\pm$0.011) & -0.112($\pm$0.005) & -38.32($\pm$1.62) \\
TYC 1877-1578-1  &2016-01-19 & F4 &324 & 6652.57($\pm$15.12)& 4.055($\pm$0.019) &  0.060($\pm$0.008) & 14.03($\pm$2.75)\\
\hline
Component& UT Date & RV & v$\sin i$ &$L_1/L_T$\\
&&(km/s)&(km/s)&\\ \hline
Primary &2015-12-20& 8.7 & 40.7& 0.938\\
        &2015-12-27& 3.6 & 78.2& 0.955\\
        &2016-01-12& 3.8 &120.0& 0.955\\
        &2016-11-04& 8.2 & 79.0& 0.955\\
\hline             
\end{tabular}
\end{table}

\section{Orbital Period Investigations} \label{sec:O-C}
\begin{figure*}
\centering
\includegraphics[width=0.8\textwidth]{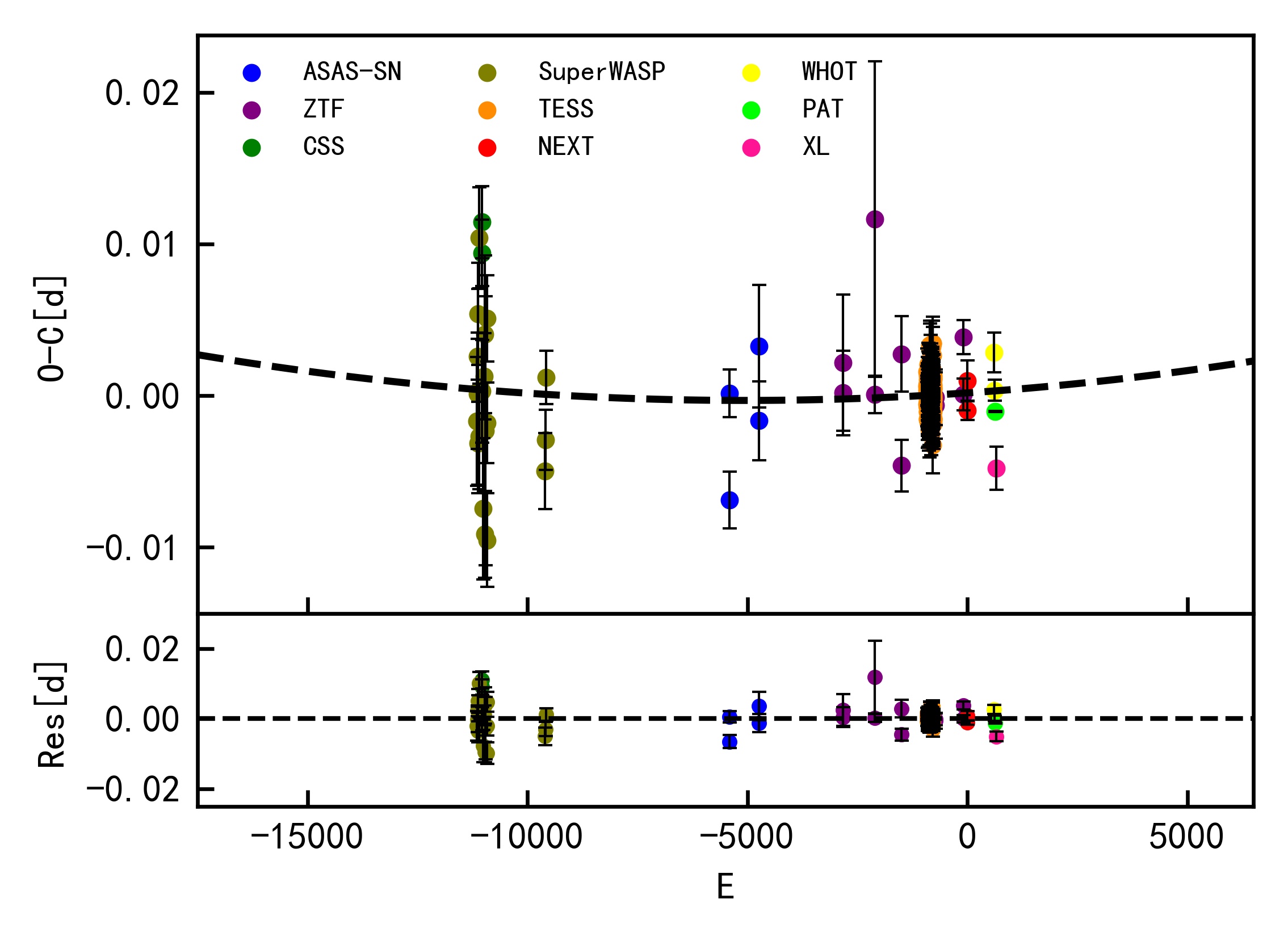}
\caption{The O-C diagram of HAT 307-0007476. The values of O-C are shown at the top of the figure, and the residuals are shown at the bottom of the figure.  \label{fig:oc}}
\end{figure*}
The study of orbital period variation is important for understanding the dynamic evolution of binary systems, the presence of companions, and material exchange between the components \citep{2016ApJ...817..133Z, 2018PASP..130g4201L, 2019ApJ...871...10U}. Therefore, research on the orbital period variation has always been crucial in binary systems. This paper presents the O-C (observed eclipsing minimum minus calculated eclipsing minimum) diagram of HAT 307-0007476. We gathered all possible eclipsing times to create O-C diagrams from various sky surveys, including the TESS \citep{2014SPIE.9143E..20R}, ASAS-SN \citep{2014ApJ...788...48S,2017PASP..129j4502K,2018MNRAS.477.3145J}, ZTF \citep{2019PASP..131a8002B, 2019PASP..131a8003M}, SuperWASP \citep{2006PASP..118.1407P, 2010A&A...520L..10B}, and the CSS \citep{2009ApJ...696..870D}.

The SuperWASP and TESS telescopes have sufficient time resolution to directly calculate eclipsing times using the K-W method \citep{1956BAN....12..327K}. We obtained 20-eclipsing times from SuperWASP data between 2006 and 2011 and 232-eclipsing times from TESS data in 2021. The other surveys are very dispersed and cannot be used directly to calculate the eclipsing times. However, we used the \cite{2020AJ....159..189L} method to group the data into suitable segments based on data quality, and the discrete data points within each group were shifted to one period, then we can calculate the eclipsing times for these surveys' data. We obtained 4-eclipsing times from ASAS-SN, 10-eclipsing times from ZTF, and 2-eclipsing times from CSS. As the observation date of TESS was given in BJD, we converted other observed minima from HJD to BJD from the online tool\footnote{\url{https://astroutils.astronomy.osu.edu/time/}}. Finally, we obtained 273-eclipsing times in total.

We first calculated the corresponding epoch and O–C values based on the Equation \ref{E0}, and we found a clear linear relationship, so we did a linear correction to get a new ephemeris:

\begin{equation}
\label{E2}
\begin{aligned}
(BJD)Min. I =&2459935.15486(\pm0.00015) + 0.5329393(\pm0.0000001)\times E.
\end{aligned}
\end{equation}
Then the corresponding epoch and O–C values were calculated based on the new Equation \ref{E2}. The O-C diagram of HAT 307-0007476 is displayed in Figure \ref{fig:oc}.

\begin{table*}\centering
\begin{center}
\caption{The Eclipsing Times of HAT 307-0007476.}\label{tab:minima}
\setlength{\tabcolsep}{5mm}{
\begin{tabular}{ccccccc}
\hline\hline
BJD & Errors & E & O-C & Residuals & Source
\\\hline

             2459935.1539 &  0.0007 &   0    &  -0.0010   &  -0.0012    & (1)  \\ 
             2459934.3564 &  0.0014 & -1.5   & 0.0010     &   0.0008    & (1) \\	
             2460256.2536 &  0.0013 &602.5   & 0.0028 	 &   0.0025    & (2) \\	
             2460267.1764 &  0.0007 &623     & 0.0003     &   0.0000    & (2)  \\
             2460271.4385 &  0.0000 &631     & -0.0011    &   -0.0014   & (3) \\	
             2460283.9588 &  0.0014 &654.5   & -0.0048    &   -0.0051   & (4)  \\
             2457053.8187 &  0.0016 &-5406.5 &  0.0002    &   0.0005    & (5)  \\
             2457054.0781 &  0.0019 &-5406   & -0.0069    &   -0.0066   & (5)  \\
             2457410.8911 &  0.0041 &-4736.5 &  0.0033    &   0.0036    & (5)  \\
             2457411.1527 &  0.0026 &-4736   & -0.0017    &  -0.0013    & (5)  \\
             2458426.9388 &  0.0045 &-2830   &  0.0022    &  0.0024     & (6)  \\
             2458427.2033 &  0.0028 &-2829.5 &  0.0002    &  0.0004     & (6) \\	
             2458814.9165 &  0.0012 &-2102   &  0.0001    &  0.0002     & (6)  \\
             2458815.1946 &  0.0104 &-2101.5 &  0.0116    &  0.0118     & (6) \\
             2454050.9814 &  0.0022 &-11041  &  0.0094     &  0.0090    & (7) \\
             2454051.2500 &  0.0024 &-11040.5 &  0.0114    &  0.0111    & (7) \\
             2453993.6794 &  0.0043 &-11148.5 &  -0.0017   &  -0.0021   & (8) \\
             2453997.6807 &  0.0016 &-11141   &  0.0026    &  0.0022    & (8) \\
             2454001.6753 &  0.0036 &-11133.5  &  0.0001  & -0.0003     & (8) \\
             2454004.6066 &  0.0067 &-11128  &  0.0003   & -0.0001      & (8) \\
             2459447.7824 &  0.0006 &-914.5  &  0.0006   & 0.0006       & (9) \\
             2459448.0490 &  0.0013 &-914  &  0.0007     & 0.0006       & (9) \\

\hline
\end{tabular}}
\end{center}
Note. (1) NEXT; (2) WHOT; (3) PAT; (4) XL; (5) ASAS-SN ; (6) ZTF ; (7) CSS; (8) SuperWASP; (9) TESS\\
(This table is available in its entirety in machine-readable form in the online version of this article.) 
\end{table*} 

As shown in Figure \ref{fig:oc}, the O-C curve shows an upward parabola trend, exhibiting a long-term orbital period increase. Using the least square method, we can obtain the following equation:
\begin{eqnarray}
\label{E:parabola}
\begin{aligned}
(BJD)Min. I =&2459935.15503(\pm0.00014) + 0.5329395(\pm0.0000001)\times E
 +1.95(\pm0.42)\times 10 ^{-11}\times E^2.
\end{aligned}
\end{eqnarray}

According to Equation \ref{E:parabola}, the period changing rate is $dP/dt= 2.67(\pm0.42)\times 10^{-8}d \cdot yr^{-1}$. This result is preliminary and needs confirmation through long-term observations due to the limited time span of eclipsing times.

\section{Discussion and Conclusions}
\label{discussion}
This paper presents the first photometric and spectroscopic investigation of HAT 307-0007476. The light curve of HAT 307-0007476 has a flat-bottomed total eclipse. We can obtain reliable physical parameters from such light curves \citep{2003CoSka..33...38P,2005Ap&SS.296..221T,2021AJ....162...13L}. We determined the mass ratio of this system to be approximately 0.114. The fill-out factor is 37.1$\%$, so it is a low mass ratio, medium contact binary \citep{2005AJ....130..224Q}. We detected two flare events in the set of light curve of December 2022 by NEXT. The statistical properties of these two flares are analyzed, including the duration, flare amplitude, energy, etc. The O-C diagram shows that this target shows a long-term orbital period increase and the period changing rate is $dP/dt=2.67(\pm0.42)\times 10^{-8}d \cdot yr^{-1}$. We found excess emissions in the H$_\alpha$ line in the LAMOST spectra of HAT 307-0007476, indicating chromospheric activity. 

%\vspace{-1em}
\subsection{Absolute parameters}\label{Absolute parameters}
Due to the unavailability of the radial velocity curve, we cannot accurately determine the absolute physical parameters of the HAT 307-0007476. However, the high precision distance provided by the Gaia DR3 \citep{2022arXiv220800211G} mission can be used to estimate the absolute physical parameters of the contact binaries without the radial velocity using the algorithm proposed by \cite{2019RAA....19..147L,2021AJ....162...13L}. First, the absolute magnitude of HAT 307-0007476 was obtained based on the following relation $M_V=m_V-5\log D+5-A_V$, where $m_V$ is the V band brightest visual magnitude which was recorded in \cite{2006AJ....131..621G}, $D=1601.6\pm44.0$ pc is the distance based on the parallax provided by Gaia DR3, $A_V=0.462$ is the V band extinction which can be obtained with an online tool called GALEXTIN\footnote{GALEXTIN (\url{http://www.galextin.org}) is an online tool to determine the interstellar extinction in the Milky Way.} \citep{2021MNRAS.508.1788A}, while choosing Baystar19 dustmap \citep{2019ApJ...887...93G}. Second, the total luminosity of the system was obtained based on the following relation, $M_{bol}=-2.5\log L/L_\odot+M_{bol\odot}$ and $M_{bol}=M_V+BC_V$, where $M_{bol}$ is the absolute bolometric magnitude, $M_{bol\odot}$ is taken as 4.73 mag \citep{2010AJ....140.1158T} and $BC_V$ is the bolometric correction which can be interpolated from Table 5 in \cite{2013ApJS..208....9P}. Third, assuming blackbody emission, the semi-major axis $a$ of this binary can be obtained according to the relation, $L=4\pi{\sigma}T_1^4({ar_1})^2+4\pi{\sigma}T_2^4({ar_2})^2$, where $r_1$ and $r_2$ are equal volume radius recorded in Table \ref{table:all solutions}, $T_\odot$ is taken as 5777K \citep{2011A&A...526A..71D,1997AJ....113.1138C}. The absolute radius of each component ($R_1$ and $R_2$) can be obtained immediately according to the relation $r=R/a$. Fourth, we calculated the luminosity of each component ($L_1$ and $L_2$) according to the averaged luminosity ratio recorded in Table \ref{table:all solutions}. Finally, we calculated the mass of each component according to Kepler's third Law $M_1+M_2=0.0134a^3/P^2$ and mass ratio $q$. Finally, the final results: $a=3.27\pm0.38R_\odot$, $M_1=1.48\pm0.52M_\odot$, $M_2=0.17\pm0.06M_\odot$, $R_1=1.90\pm0.22R_\odot$, $R_2=0.74\pm0.09R_\odot$, $L_1=7.11\pm0.39L_\odot$, $L_2=0.89\pm0.05L_\odot$.

\subsection{Secular orbital period increase} \label{period analyse}
Usually, an increase in the orbital period over a long time is due to mass transfer from the less massive component to the more massive component. Using the following equation \citep{1958BAN....14..131K}: 
\begin{eqnarray}
\frac{\dot{P}}{P} = -3\dot{M_1}(\frac{1}{M_1}-\frac{1}{M_2}),
\end{eqnarray}
we determine the mass transfer rate $dM_1/dt = 3.00(\pm0.59)\times 10^{-9}M_\odot yr^{-1}$. The positive value means the more massive primary component is receiving mass. 

\subsection{The Flare Events} \label{sec:Solutions}
Flare activity is detected on HAT 307-0007476. We analyzed these flares' statistical properties. A total of two flare events have been detected on this target observed by NEXT (see Table \ref{table:flares}). The interval between the two flares is 4 days, they are highly randomized in association with our observing schedule, so we did not calculate the flare frequency because the results would have been over-estimated. We discuss the statistical properties of these flares, including the durations, amplitudes, equivalent width, energies, and the relationship between flare and orbital phase.
\subsubsection{Flare Duration, Amplitude, Equivalent width, and Energy}\label{sec:frequency}
Flares are typically characterized by several parameters, including amplitude, duration, equivalent width (EW), and energy. We will define and calculate these eigenvalues using the following calculation method. We can see that all the flares in Figure \ref{fig:flares} have a rapid rise followed by a slowing exponential decay. We take the time to rise to maximum luminosity as the rise time and the time to recover the intrinsic luminosity from the maximum luminosity as the decay time. The duration is expressed as the sum of the rise time and the decay time. The average duration of Flare1 is about 2083 seconds, and the average duration of Flare2 is about 2496 seconds. The amplitude is a parameter that indicates the strength of a flare. The amplitude was calculated by comparing the observed and theoretical light curves with no flares. The theoretical light curve is calculated using the W-D code. It can be calculated as
\begin{equation}
\label{E:Amp}
\begin{aligned}
Amplitude = M_{peak} - M_{thero},
\end{aligned}
\end{equation}
where $M_{peak}$ is the magnitude of the flare peak in the observed curve, and $M_{thero}$ is the magnitude in the theoretical light curve. We calculated the amplitudes of the flares in each band and listed them in Table \ref{table:flares}.

\begin{center}
\begin{table}\centering
\footnotesize
\caption{Parameters of Observed Flares for the HAT 307-0007476}
\label{table:flares}
\setlength{\tabcolsep}{2mm}{
\begin{tabular}{lccccccccc}
\hline\hline
Number & Time & Filter & Phase & Amplitude & Duration & $T_{rise}$ & $T_{decay}$  &EW& $E_{flare}$ \\
&&&&(mag)&(s)&(s)&(s)&(s)&(erg)
\\\hline
	
Flare1 &2022.12.21 &$g^{\prime}$ &0.854 &0.042 &2083 &779 &1304 & &    \\
       &           &$r^{\prime}$ &0.850 &0.037 &1823 &519 &1304 &    \\
       &           &$i^{\prime}$ &0.858 &0.061 &2345 &1041 &1304 &    \\
       &           &  & & & & & &91 &2.8e+36    \\
Flare2 &2022.12.25 &$g^{\prime}$ &0.494 &0.041  &2061 &750 &1311 &    \\
       &           &$r^{\prime}$ &0.501 &0.027  &2583 &1012 &1571 &    \\
       &           &$i^{\prime}$ &0.497 &0.038  &2843 &1010 &1833 &    \\
       &           &  & & & & &  &22 &6.8e+35    \\
%Flare3  &2021.8.30 &TESS&0.457&1.823 & 17400&7200&10200&16514&5.1e+38 \\
%Flare4 &2021.8.31  &TESS&0.825&1.338 & 15600&3793&11807&2855 &8.8e+37  \\
\hline
\end{tabular}}
\end{table}
\end{center}

We calculated the EWs of all these flares with Equation \ref{E3}. For each flare, we integrate the points that are tagged as part of the flare in the curve of the observed curve minus the theoretical curve, which is essentially a photometric EW of the flare \citep{2011AJ....141...50W}, i.e.,
\begin{eqnarray}
\label{E3}
EW=\int \frac{F_{\rm f}-F_{\rm q}}{F_{\rm q}} dt,
\end{eqnarray}
where $F_{\rm f}$ and $F_{\rm q}$ are the flaring and the quiescent flux, respectively. Finally, the EW of Flare1 is $\sim$ 91s and the EW of Flare2 is $\sim$ 22s. The bolometric energy of each flare using the following equation: $E_{flare}= L_{\ast}\int \frac{F_{\rm f}-F_{\rm q}}{F_{\rm q}} dt (erg)$, where $L_{\ast}$ is the stellar luminosity \citep{2013ApJS..209....5S, 2015ApJ...798...92W, 2017ApJ...834...92C, 2018ApJ...867...78C, 2019ApJ...873...97L} calculated in Section \ref{Absolute parameters}. The values of all these flares' energy are presented in the last column of Table \ref{table:flares}. The energy of Flare1 is about $3\times 10^{36}$ erg and the energy of Flare2 is about $7\times 10^{35}$ erg. These two flares are more than 100 times greater than the largest white-light solar flare, reaching the superflare level \citep{2000ApJ...529.1026S,2011AJ....141...50W, 2014ApJ...792...67C,2014ApJ...797..121H,2017ApJ...834...92C}.

We also study the orbital phase in which the flare is observed (see Table \ref{table:flares}). The Flare2 occurred near phase $\sim0.5$, at which most luminosity comes from the primary component. Flare1 occurred near phase $\sim0.8$, where the secondary component was egress. It's difficult to determine which component the flare is coming from just from the current two flare phases.

\subsubsection{Relations between Flare Energy, Flare Duration, and
Amplitude}

\begin{figure*}
\begin{minipage}[t]{0.49\textwidth}
\includegraphics[width=\textwidth, angle=0]{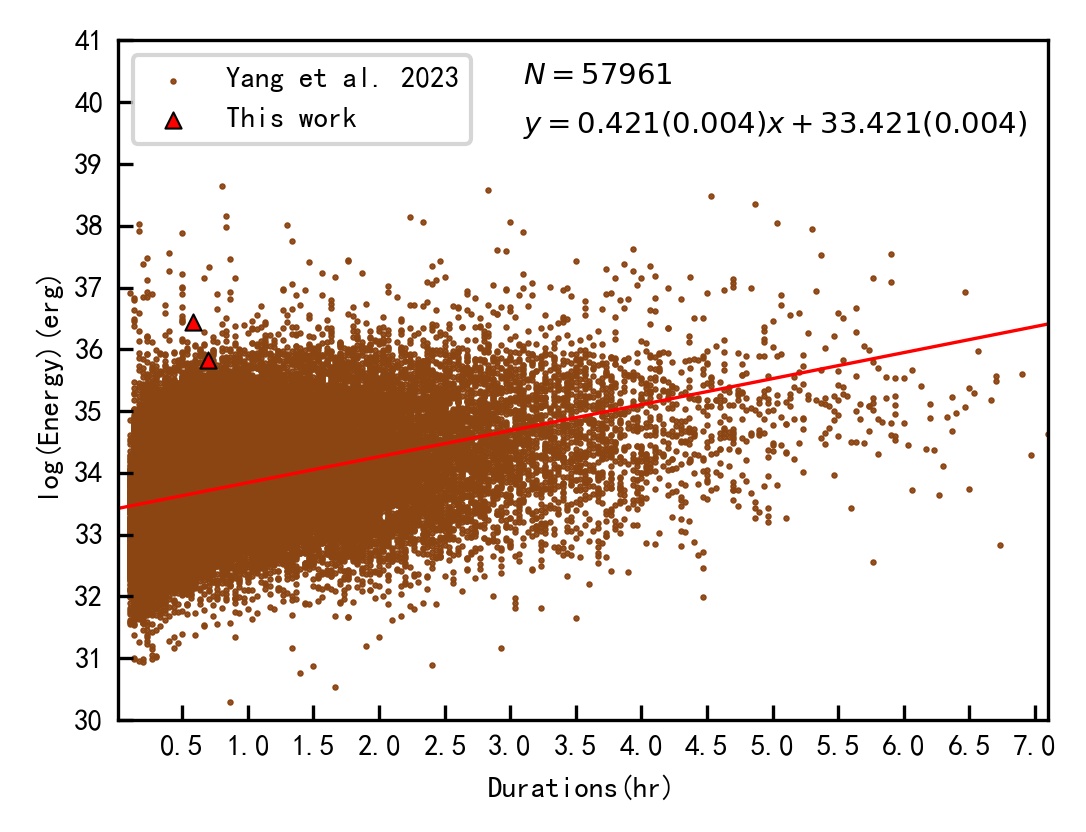}
\end{minipage}
\begin{minipage}[t]{0.49\textwidth}
\includegraphics[width=\textwidth, angle=0]{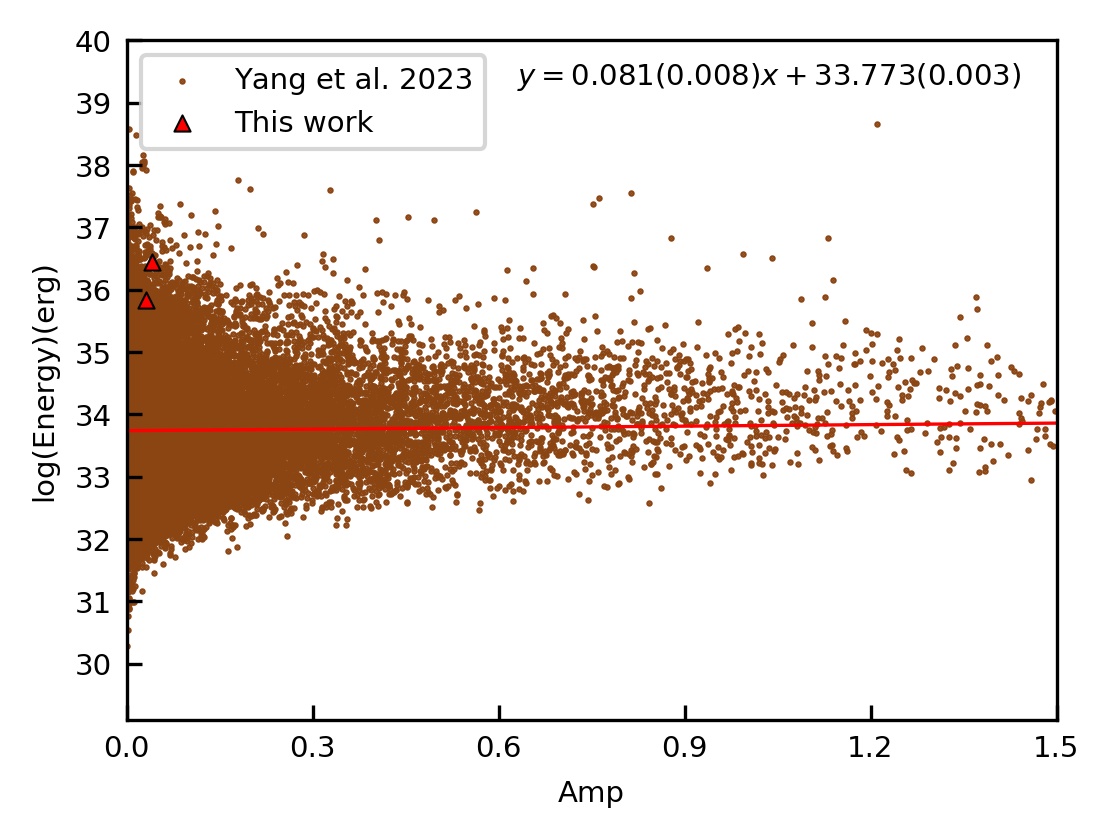}\hspace{-5cm}
\end{minipage}
\caption{Relationships between flare energy and duration (left) or amplitude (right). The red triangles represent flares from this work. The brown dots represent flares from Y23 and the solid lines are linear fits from Y23. \label{fig:energy} }
\end{figure*}
\cite{2023A&A...669A..15Y} (hereafter Y23) used 2-min cadence light curves in the first 30 sectors of the TESS data from July 2018 to October 2020 and detected 60810 flare events on 13478 stars. Both Y23 and our work have the same data source from TESS, and the sample of Y23 size is large enough to get a more comprehensive understanding of the relations between flare energy, flare duration, and amplitude. Therefore, we marked our two flare events based on Figure 13 of Y23 (see Figure \ref{fig:energy}).

Figure \ref{fig:energy} represents the relationships between flare energy and duration (left) or amplitude (right). The red triangles represent flares from our work. The brown dots represent flares from Y23 and the solid lines are linear fits from Y23. There is a positive correlation between flare energy and flare duration. The longer the duration of the flare, the stronger the energy and the larger the amplitude of the flare. As seen in Figure \ref{fig:energy}, we found that corresponding to the same duration or amplitude, the energy of these two flare events detected on this object belong to the upper side, and we suggest that the possible reason is that the spectral type of this binary is A9/F0, with a total system luminosity of about 8$L_\odot$, which leads to a higher flare energy. Another reason is that the sample of the A/F-type flare stars in Y23 accounts for less than 10$\%$ of the total sample, and the frequency of the flare of the A/F-type stars is lower than G/K/M type stars.

\subsection{Instability Parameter}
HAT 307-0007476 is a low mass ratio contact binary. The ratio of the spin angular momentum ($J_{spin}$) to the orbital angular momentum ($J_{orb}$) is very important to examine the dynamic stability of contact binaries \citep{1980A&A....92..167H}. The contact binary would be instability (Darwin's instability) when $J_{spin} \geqslant \frac{1}{3} J_{orb}$. To determine whether the HAT 307-0007476 is in the stable region, we calculated the $J_{spin}/J_{orb}$ according to the equation \citep{2015AJ....150...69Y},
\begin{equation}
\begin{aligned}
\dfrac{J_{spin}}{J_{orb}}=\dfrac{1+q}{q}[(k_{1}r_{1})^{2}+q((k_{1}r_{1})^{2})]\\
\end{aligned}
\end{equation}
where $q$ is the mass ratio, $r_{1}$, $r_{2}$ is the relative radius, and $k_{1}$, $k_{2}$ is the dimensionless rotation radius. The values of $k_{1}^{2}=k_{2}^{2}=0.06$ were adopted from \cite{2006MNRAS.369.2001L}. We found the $J_{spin}/J_{orb}$ of HAT 307-0007476 was 0.21. $J_{spin}/J_{orb}$ of of the HAT 307-0007476 is less than $\frac{1}{3}$, so it is dynamically stable. In fact, the value of $k_{1}$, $k_{2}$ can be further precise. Because the mass of the secondary is very low, it should be a fully convective star, $k_2^{2}=0.205$ was adopted \citep{2007MNRAS.377.1635A}. We reestimated $k_{1}$ for primary component using the tabulated results from \cite{2009A&A...494..209L}, which considers the effects of tidal and rotational distortions on a star in binary system. We calculated $k_{1}$ using the following linear relationships, $k_{1}=0.014M+0.152$ ($M_1 \textgreater 1.4M_{\odot}$). After re-determining $k_{1}$, $k_{2}$, we calculated $J_{spin}/J_{orb}$ again and the result is 0.11. The values of $J_{spin}/J_{orb}$ are all listed in Table \ref{table6}.
\begin{table}
\begin{center}
\caption{The instability parameters of HAT 307-0007476.}
\setlength{\tabcolsep}{4mm}{
\label{table6}
\begin{tabular}{ccccccccc} 
\hline
$J_{spin}/J_{orb}$   & $J_{spin}/J_{orb}$   &$q_{inst}$& $A_{inst}$ ($R_{\odot}$)& $P_{inst} (d)$\\  
        $(k_{1}=k_{2})$ & $(k_{1}\neq k_{2})$ &      &                         &       \\
\hline
 0.21           & 0.11            & 0.0529   & 2.265                   & 0.3489\\

\hline
\end{tabular}}
\end{center}
\end{table}

Recently, \cite{2021MNRAS.501..229W} proposed the instability mass ratio, the instability separation, and the instability period of contact binaries. We calculated the instability parameters of HAT 307-0007476 following \cite{2021MNRAS.501..229W}, with the stability criterion of $dJ_{T}/dJ_{A}= 0$. For stars of different masses, there exists a linear relationship between $f$ and $q_{inst}$. Using Equation (14), (15) in \cite{2021MNRAS.501..229W}, the instability mass ratio can be determined for any given fill-out factors. The instability separation can be obtained by Equations (7), (8), (10) in \cite{2021MNRAS.501..229W}. After getting the instability separation, the instability period can be calculated by Kepler's third law. These instability parameters of HAT 307-0007476 are all listed in Table \ref{table6}. The instability parameters are smaller compared to the current values, indicating that the HAT 307-0007476 is relatively stable.

In conclusion, we reported an interesting long-period total eclipsing contact binary (HAT 307-0007476) in this work. Photometric and spectroscopic investigations of this target were carried out for the first time. Two flare events were detected in multiple bands in December 2022. The average duration of these two flares is about 2289s. Both the two flares reach superflare energy levels. This system is a low mass ratio ($q\sim0.114$) and medium contact binary ($f\sim37.1\%$). The excess emission of the H$_\alpha$ line in the LAMOST spectra of this object was analyzed. The $O-C$ diagram showed a long-term orbital period increase, and the mass transfer was used to explain this long-term orbital period increase. We conclude that HAT 307-0007476 is currently stable based on both $J_{spin}/J_{orb}$ and the comparison between the instability parameters and its current values.

\section*{Acknowledgements}

The authors thank the reviewer for the helpful suggestions and comments that have contributed significantly to the improvement of this manuscript. This work is supported by National Natural Science Foundation of China (NSFC) (No.12273018), and the Joint Research Fund in Astronomy (No. U1931103) under cooperative agreement between NSFC and Chinese Academy of Sciences (CAS), and by the Natural Science Foundation of Shandong Province (Nos. ZR2014AQ019), and by the Qilu Young Researcher Project of Shandong University, and by the Young Data Scientist Program of the China National Astronomical Data Center and by the Cultivation Project for LAMOST Scientific Payoff and Research Achievement of CAMS-CAS, and by the Chinese Academy of Science Interdisciplinary Innovation Team, and by Tianshan Innovation Team Program of Xinjiang Uygur Autonomous Region (No. 2024D14015), and by Natural Science Foundation of Xinjiang Uygur Autonomous Region (No. 2022D01A361). The calculations in this work were carried out at Supercomputing Center of Shandong University, Weihai.
	
The spectral data were provided by Guoshoujing Telescope (the Large Sky Area Multi-Object Fiber Spectroscopic Telescope LAMOST), which is a National Major Scientific Project built by the Chinese Academy of Sciences. Funding for the project has been provided by the National Development and Reform Commission. LAMOST is operated and managed by the National Astronomical Observatories, Chinese Academy of Sciences.

We acknowledge the support of the staff of the Xinglong 85cm
telescope, NEXT, PAT, and WHOT. This work was partially supported by the Open Project Program of the Key Laboratory of Optical Astronomy, National Astronomical Observatories, Chinese Academy of Sciences, and by the cooperation program of Xinjiang Astronomical Observatory, Chinese Academy of Sciences' Nanshan Observatory.

This work includes data collected by the TESS mission. Funding for the TESS mission is provided by NASA Science Mission Directorate. We acknowledge the TESS team for its support of this work.

This paper makes use of data from ASAS-SN. ASAS-SN is funded in part by the Gordon and Betty Moore Foundation through grants GBMF5490 and GBMF10501 to the Ohio State University, and also funded in part by the Alfred P. Sloan Foundation grant G-2021-14192.

This paper makes use of data from CSS. The CSS is funded by NASA under grant no.NNG05GF22G issued through the Science Mission Directorate Near-Earth Objects Observations Program. The Catalina Real-Time Transient Survey is supported by the U.S. National Science Foundation (NSF) under grant nos. AST0909182 and AST-1313422.

This work has made use of data from the European Space Agency (ESA) mission Gaia (\url{https://www.cosmos.esa.int /gaia} ), processed by the Gaia Data Processing and Analysis Consortium (DPAC, \url{https:// www.cosmos.esa.int/web/gaia/dpac/consortium}). Funding for the DPAC has been provided by national institutions, in particular the institutions participating in the Gaia Multilateral Agreement.

%%%%%%%%%%%%%%%%%%%%%%%%%%%%%%%%%%%%%%%%%%%%%%%%%%
\section*{Data Availability}
The TESS data are publicly available at \href{http://archive.stsci.edu/tess/bulk_downloads.html}{http://archive.stsci.edu/tess/bulk downloads.html}. The NEXT/WHOT/XL/PAT data used in this research will be
shared on reasonable request to the corresponding author.

%%%%%%%%%%%%%%%%%%%% REFERENCES %%%%%%%%%%%%%%%%%%

% The best way to enter references is to use BibTeX:

\bibliographystyle{mnras}
\bibliography{J03} % if your bibtex file is called example.bib

% Alternatively you could enter them by hand, like this:
% This method is tedious and prone to error if you have lots of references
%\begin{thebibliography}{99}
%\bibitem[\protect\citeauthoryear{Author}{2012}]{Author2012}
%Author A.~N., 2013, Journal of Improbable Astronomy, 1, 1
%\bibitem[\protect\citeauthoryear{Others}{2013}]{Others2013}
%Others S., 2012, Journal of Interesting Stuff, 17, 198
%\end{thebibliography}

%%%%%%%%%%%%%%%%%%%%%%%%%%%%%%%%%%%%%%%%%%%%%%%%%%

% Don't change these lines
\bsp	% typesetting comment
\label{lastpage}
\end{document}